\documentclass[%
 reprint,
 superscriptaddress,
 amsmath,amssymb,
 aps,
 prl
]{revtex4-2}
\usepackage[T1]{fontenc}
\usepackage{graphicx}
\usepackage{dcolumn}
\usepackage{bm}
\usepackage[colorlinks=true, allcolors=blue]{hyperref}
\usepackage{braket}
\usepackage[table]{xcolor}
\usepackage{xfrac}
\usepackage{comment}
\usepackage[normalem]{ulem}
\usepackage{mathtools}
\usepackage{layouts}
\usepackage{bbm}
\usepackage[section]{placeins}

\usepackage{tabularx}   
\usepackage{booktabs}   
\usepackage{longtable}  
\usepackage{multirow}

\usepackage{algpseudocode}

\usepackage{quantikz}
\usepackage{tabularx}
\usepackage{multirow}
\usepackage{esint}
\usepackage{enumitem}
\usetikzlibrary{arrows.meta}
\definecolor{c1}{RGB}{178,34,34}
\definecolor{c2}{RGB}{31,90,166}
\definecolor{c3}{RGB}{34,139,34}
\definecolor{c4}{RGB}{218,145,0}

\usepackage{amsthm}
\usepackage[capitalise]{cleveref}
\newcommand{\ind}{\mathbbm{1}}

\newcommand{\supp}{\operatorname{supp}}

\newcommand{\approptoinn}[2]{\mathrel{\vcenter{
  \offinterlineskip\halign{\hfil$##$\cr
    #1\propto\cr\noalign{\kern2pt}#1\sim\cr\noalign{\kern-2pt}}}}}

\providecommand{\boldsymbol}[1]{\boldsymbol{#1}}

\definecolor{xcol}{RGB}{31,90,166}
\definecolor{zcol}{RGB}{178,34,34}
\begin{document}
\title{Disassembling qLDPC codes for depth-optimal parity-check circuits}
\author{Minh T. P. Nguyen}
\email{m.t.phamnguyen@tudelft.nl}
\affiliation{QuTech and Kavli Institute of Nanoscience, Delft University of Technology, Lorentzweg 1, 2628 CJ Delft, The Netherlands}
 \author{Maximilian Rimbach-Russ}
 \affiliation{QuTech and Kavli Institute of Nanoscience, Delft University of Technology, Lorentzweg 1, 2628 CJ Delft, The Netherlands}
\author{Stefano Bosco}
\email{s.bosco@tudelft.nl}
\affiliation{QuTech and Kavli Institute of Nanoscience, Delft University of Technology, Lorentzweg 1, 2628 CJ Delft, The Netherlands}

\newtheorem{theorem}{Theorem}[section]
\newtheorem{observation}{Observation}
\newtheorem{corollary}[theorem]{Corollary}
\newtheorem{lemma}[theorem]{Lemma}
\newtheorem{proposition}[theorem]{Proposition}
\newtheorem{criterion}{Criterion}
\newtheorem{appendixcriterion}{Criterion}[section]
\theoremstyle{definition}
\newtheorem{assumption}{Assumption}
\theoremstyle{remark}
\newtheorem*{remark}{Remark}
\theoremstyle{definition}
\newtheorem{definition}[theorem]{Definition}
\Crefname{criterion}{Criterion}{Criteria}
\crefname{criterion}{Crit.}{Crit.}

\makeatletter
\newcounter{algbox}
\newcommand{\algbox@top}[1]{%
  \refstepcounter{algbox}%
  \par\noindent\rule{\linewidth}{0.9pt}\par\vspace{2.5pt}%
  \noindent\textbf{Box~\thealgbox}\quad #1\par\vspace{2.5pt}%
  \noindent\rule{\linewidth}{0.4pt}\par\vspace{2pt}}
\newcommand{\algbox@bot}{\par\vspace{2pt}\noindent\rule{\linewidth}{0.9pt}\par}
\newenvironment{algbox}[2][t]{\begin{figure}[#1]\algbox@top{#2}}%
  {\algbox@bot\end{figure}}
\newenvironment{algbox*}[2][t]{\begin{figure*}[#1]\algbox@top{#2}}%
  {\algbox@bot\end{figure*}}
\makeatother

\begin{abstract}
Quantum low-density parity-check (qLDPC) codes offer a promising route to scalable fault-tolerant quantum computing, but their practical implementation requires efficient circuits for syndrome extraction. Many qLDPC families are assembled from a small set of components through explicit operations that imprint edge symmetries on their Tanner graphs. 
We show that disassembling the code by quotienting its symmetries one at a time, reduces parity-check circuit design to a much smaller problem acting only on the underlying components. Solutions to the reduced problem can then be lifted to the full code, yielding circuit constructions that apply to entire families of codes built from the same components. For lifted-product codes, our approach provides an optimal analytical construction for parity-check circuits achieving the lower bound on CNOT depth. We obtain analogous constructions for balanced-product codes and  quantum--classical homological product codes. For quantum Tanner codes, the reduced problem is small enough to solve numerically, reaching the CNOT-depth lower bound for nearly all instance tested, including codes with up to 576 data qubits.
\end{abstract}

\maketitle

\paragraph*{Introduction.--}
A major advance in quantum error correction has been the discovery of good quantum low-density parity-check (qLDPC) codes~\cite{Breuckmann_2021} that have high encoding rate and relative code distance, opening a promising path toward scalable fault-tolerant quantum computation~\cite{gottesman2014faulttolerantquantumcomputationconstant,cain2026shorsalgorithmpossible10000,yoder2025tourgrossmodularquantum}. Prominent families of qLDPC codes include hypergraph-product~\cite{Tillich_2014}, fiber-bundle~\cite{MatthewFiberBundle2021}, lifted-product ~\cite{Panteleev_2021,Panteleev_2022,panteleev2022asymptoticallygoodquantumlocally}, balanced-product~\cite{BreuckmannBalancedProduct_2021}, and quantum Tanner codes~\cite{leverrier2022quantumtannercodes}. Beyond code constructions, substantial progress has been made toward realizing their full computational potential, including logical operations~\cite{Cohen_2022,Cowtan_2024,Williamson_2026,swaroop2025universaladaptersquantumldpc}, efficient decoders~\cite{Panteleev_2021,Roffe_2020,ye2025beamsearchdecoderquantum,koutsioumpas2025automorphismensembledecodingquantum}, and reduced hardware requirements~\cite{McEwen_2023,Shaw_2025}.

A comparatively less studied, but equally critical, step is syndrome extraction, typically performed by parity-check circuits. The checks can always be measured sequentially, all $X$-checks before all $Z$-checks, but this alternating circuit has high CNOT depth and leaves the data qubits idling. These limitations motivate \emph{interleaved} parity-check circuits, where $X$- and $Z$-checks share CNOT layers. To our knowledge no general analytic construction of interleaved circuits for Calderbank-Shor-Steane (CSS) codes is known; instead they have been designed case by case for specific families \cite{Tomita_2014,Bravyi_2024,Menon_2026}. Recent works automate the task with differing objectives: Ref.~\cite{zhang2026optimalsyndrome} targets the shortest interleaved circuit, while Refs.~\cite{liu2026alphasyndrome,viszlai2026prophunt,strikis2026syndromeextractioncircuits} optimize performance under circuit-level noise or specific hook-error patterns, generally at the cost of longer circuits. Greedy constructions~\cite{Tremblay2022,vittal2024flagproxynetworkstackling,Kang_2025} typically achieve low CNOT depth, but not the theoretical minimum. Alternative approaches, e.g. amortized CNOT depth~\cite{strikis2026syndromeextractioncircuits} and morphing circuits~\cite{McEwen_2023,Shaw_2025,shaw2026optimisingquantumerrorcorrection} can bring additional benefits, e.g. reduced hardware connectivity requirements.

\begin{figure}
    \centering
    \begin{tikzpicture}[
  bit/.style={circle,draw,fill=white,inner sep=1.2pt},
  chk/.style={rectangle,draw,fill=black!75,inner sep=1.5pt},
  lab/.style={font=\scriptsize},
  arr/.style={-{Stealth[length=2mm]},thick},
  rarr/.style={{Stealth[length=2mm]}-,thick,gray!95},
  blob/.style={gray!50,fill=gray!7,rounded corners=4pt}]
\begin{scope}[shift={(0,-0.31)}]
\draw[blob] (-0.22,-0.3) rectangle (1.22,0.92);
\node[chk] (a1) at (0.1,0.3) {}; \node[bit] (a2) at (0.55,0.62) {};
\node[bit] (a3) at (0.6,0.02) {}; \node[chk] (a4) at (1.0,0.34) {};
\draw[c1,very thick] (a1)--(a2); \draw[c2,very thick] (a1)--(a3);
\draw[c3,very thick] (a4)--(a2); \draw[c4,very thick] (a4)--(a3);
\node[lab] at (-0.5,0.3) {$T_0$};
\end{scope}
\begin{scope}[shift={(2.5,-0.6)}]
\draw[blob] (-0.25,-0.25) rectangle (1.32,1.45);
\node[bit] (p1) at (0.0,0.25) {}; \node[chk] (p2) at (0.42,0.0) {};
\node[bit] (p3) at (0.92,0.12) {}; \node[chk] (p4) at (0.12,0.78) {};
\node[bit] (p5) at (0.6,0.6) {};  \node[chk] (p6) at (1.07,0.65) {};
\node[bit] (p7) at (0.35,1.18) {};\node[chk] (p8) at (0.9,1.2) {};
\draw[c1,very thick] (p1)--(p2); \draw[c2,very thick] (p2)--(p3);
\draw[c3,very thick] (p1)--(p4); \draw[c1,very thick] (p4)--(p5);
\draw[c4,very thick] (p5)--(p2); \draw[c2,very thick] (p5)--(p6);
\draw[c3,very thick] (p3)--(p6); \draw[c2,very thick] (p4)--(p7);
\draw[c1,very thick] (p7)--(p8); \draw[c3,very thick] (p8)--(p5);
\node[lab] at (0.53,1.72) {$T_1$};
\end{scope}
\begin{scope}[shift={(5.85,-0.9)}]
\draw[blob] (-0.25,-0.25) rectangle (2.12,2.05);
\node[bit] (e1) at (0.05,0.15) {}; \node[chk] (e2) at (0.5,0.0) {};
\node[bit] (e3) at (1.0,0.1) {};   \node[chk] (e4) at (1.5,0.05) {};
\node[chk] (e5) at (0.2,0.62) {};  \node[bit] (e6) at (0.72,0.55) {};
\node[chk] (e7) at (1.27,0.6) {};  \node[bit] (e8) at (1.8,0.5) {};
\node[bit] (e9) at (0.05,1.18) {}; \node[chk] (e10) at (0.6,1.12) {};
\node[bit] (e11) at (1.1,1.22) {}; \node[chk] (e12) at (1.62,1.3) {};
\node[bit] (e13) at (0.85,1.72) {};
\draw[c1,thick] (e1)--(e2); \draw[c2,thick] (e2)--(e3); \draw[c1,thick] (e3)--(e4);
\draw[c3,thick] (e4)--(e8); \draw[c2,thick] (e1)--(e5); \draw[c1,thick] (e5)--(e6);
\draw[c2,thick] (e6)--(e7); \draw[c1,thick] (e7)--(e8); \draw[c3,thick] (e6)--(e2);
\draw[c3,thick] (e5)--(e9); \draw[c1,thick] (e9)--(e10); \draw[c2,thick] (e10)--(e11);
\draw[c1,thick] (e11)--(e12); \draw[c2,thick] (e12)--(e8); \draw[c3,thick] (e10)--(e13);
\draw[c4,thick] (e13)--(e12); \draw[c4,thick] (e7)--(e11); \draw[c4,thick] (e3)--(e10);
\node[lab] at (0.93,2.32) {$T_{\mathcal{C}}$};
\end{scope}
\draw[arr]  (1.32,0.22) -- (2.18,0.22) node[midway,above,lab] {Op.\,1};
\draw[rarr] (1.32,-0.22) -- (2.18,-0.22) node[midway,below,lab] {$\mathcal P_1$};
\draw[arr]  (3.95,0.22) -- (4.45,0.22) node[midway,above,lab] {Op.\,2};
\draw[rarr] (3.95,-0.22) -- (4.45,-0.22) node[midway,below,lab] {$\mathcal P_{2}$};
\node[font=\normalsize] at (4.75,0) {$\cdots$};
\draw[arr]  (5.05,0.22) -- (5.55,0.22) node[midway,above,lab] {Op.\,$N$};
\draw[rarr] (5.05,-0.22) -- (5.55,-0.22) node[midway,below,lab] {$\mathcal P_N$};
\end{tikzpicture}
    \caption{\textbf{Assembling and disassembling codes.} A Tanner graph $T_0$ is assembled into a qLDPC code $T_{\mathcal{C}}$ through a sequence of operations (black arrows). Rather than constructing the parity-check circuit directly on $T_{\mathcal{C}}$, we reverse these operations by quotienting the graph edge symmetries (gray arrows), solve the reduced parity-check scheduling problem (edge colors) on $T_0$, and lift the resulting circuit back to $T_{\mathcal{C}}$.}
    \label{fig: growing diagram for qLPDC code}
\end{figure}

In this work, we introduce a strategy for constructing low-CNOT-depth interleaved parity-check circuits for CSS qLDPC codes. Our key observation is that many good qLDPC codes \cite{BreuckmannBalancedProduct_2021,MatthewFiberBundle2021,panteleev2022asymptoticallygoodquantumlocally,Breuckmann_2021} are assembled from a small set of components through explicit operations, such as hypergraph products or group lifts, that imprint edge symmetries on their Tanner graphs~\cite{rakovszky2024physicsgoodldpccodes}. To obtain a reduced problem, we disassemble the code by quotienting these symmetries and construct the parity-check circuit on the reduced and significantly simpler graph. We then lift the solution to the full code, as sketched in Fig.~\ref{fig: growing diagram for qLPDC code}. For Lifted Product (LP) codes, this gives an analytical parity-check circuit within one layer of the lower bound on CNOT depth. This construction can be extend to balanced product codes and to homological products of quantum and classical codes. For Quantum Tanner (QT) codes, we use the structure of the reduced problem to devise a simple and fast greedy algorithm that reaches the lower bound on nearly every instance we test.

\paragraph*{Scheduling parity-checks.--}
We consider a general CSS qLDPC code $\mathcal{C}$ with parity-check matrices $H_X$ and $H_Z$. Its Tanner graph $T_{\mathcal C}=(V,E)$ has vertices $V=V_Q\cup V_X\cup V_Z$ grouping data ($Q$) and ancilla qubits for $X$- and $Z$-checks, and edges $E=E_X\cup E_Z$ for $X$- and $Z$-checks. When unambiguous, we identify a check vertex $v_x\in V_X$ (resp.\ $v_z\in V_Z$) with the row of $H_X$ (resp.\ $H_Z$) it measures, so that $\supp(v_x)\subseteq V_Q$ are the data qubits in its support.

A parity-check circuit extracts the syndrome by preparing each $X$- ($Z$-) ancilla in $\ket{+}$ ($\ket{0}$) and applying layers of CNOT gates. We require that each data or ancilla qubit participates in at most one gate per layer and impose no restriction on qubit connectivity. Designing such a circuit can naturally be formulated as a job-scheduling problem \cite{zhang2026optimalsyndrome}. Specifically, we assign to each edge $e\in E$ a time label $\tau(e)\in\mathbb Z_{+}$ specifying the CNOT layer when the corresponding gate is applied. A parity-check circuit is then specified as the set of time labels $\{\tau(e)\}$, with CNOT depth given by $\max_{e\in E}\tau(e)$.

Two restrictions on $\{\tau(e)\}$ are required for a valid parity-check circuit. First, because a qubit can take part in only one gate per layer, edges sharing a vertex must carry distinct time labels, i.e.
\begin{equation}
\label{eq: incidence edge label constraint}
\tau(e_1)\neq\tau(e_2)\quad\text{ if } e_1\cap e_2\neq\emptyset \ .
\end{equation}
Second, an $X$-operator on the ancilla $v_x$ should propagate only to the data qubits in $\supp(v_x)$, $X_{v_x}\rightarrow X_{v_x}\otimes\supp(v_x)$, and analogously for a $Z$-operator on $v_z$. We call this property the \textit{properness} of the circuit \cite{Conrad_2018}. An improper scheduling of CNOT gates on data qubits shared by $v_x$ and $v_z$ can  lead to contamination, i.e. $X_{v_x}\rightarrow X_{v_x}\otimes\supp(v_x)\otimes X_{v_z}$. Contamination is a critical issue as $X_{v_z}$ randomizes the measurement of the $v_z$ ancilla even in noiseless circuits. Improper parity-check circuits can be corrected by appending CNOT gates between ancillas, at the cost of increasing the CNOT depth and requiring higher connectivity in hardware. 

Whether a pair of checks is proper depends only on the ordering of the CNOT gates on their shared data qubits. As shown in the Supplementary Material (SM), a parity-check circuit is proper on a pair of checks iff the $X$-check interacts with the shared data qubits before the $Z$-check an even number of times, i.e.~\cite{Conrad_2018, zhang2026optimalsyndrome}
\begin{equation}
\label{eq: cpsat constraint}
\smashoperator[r]{\sum_{v_q\in\supp(v_x)\cap\supp(v_z)}}
\ind\bigl[\tau(v_x-v_q)<\tau(v_z-v_q)\bigr]\equiv 0
\!\!\!\pmod 2 \ ,
\end{equation}
with $\ind[a<b]$ being the indicator function of $a<b$. We show in the SM that our parity-check circuits automatically preserve stabilizers and logical operators, so we focus on enforcing Eqs.~\eqref{eq: incidence edge label constraint} and~\eqref{eq: cpsat constraint}.

If we neglect Eq.~\eqref{eq: cpsat constraint}, minimizing the CNOT depth reduces to an edge-coloring problem on $T_{\mathcal C}$, which by K\"onig's theorem \cite{cook2011combinatorial} requires exactly $\Delta=\deg(T_{\mathcal C})$ colors, each corresponding to a CNOT layer. In contrast, the alternating parity-check circuit is always valid and has depth $\Delta_X+\Delta_Z$, where $\Delta_X$ and $\Delta_Z$ are the maximum degrees of $T_{\mathcal C}$ restricted to the $X$- and $Z$-checks, respectively. Hence, the optimal CNOT depth satisfies
\begin{equation}
\Delta \;\le\; \min_{\tau}\max_{e \in E} \tau(e) \;\le\; \Delta_X+\Delta_Z \ .
\end{equation}

\begin{figure}[htb!]
\centering
\hspace*{-4mm}%
\begin{tikzpicture}[trim left=-4.4cm, trim right=4.9cm, >=Stealth, scale=0.85, transform shape,
  data/.style={circle,draw,line width=0.8pt,fill=blue!8,minimum size=13mm,inner sep=1pt,
               align=center,font=\scriptsize},
  xchk/.style={rectangle,rounded corners=1pt,draw,line width=0.8pt,fill=red!12,
               minimum width=13mm,minimum height=10mm,inner sep=1pt,align=center,font=\scriptsize},
  zchk/.style={rectangle,rounded corners=1pt,draw,line width=0.8pt,fill=green!14,
               minimum width=13mm,minimum height=10mm,inner sep=1pt,align=center,font=\scriptsize},
  fbit/.style={circle,draw,fill=blue!8,minimum size=6mm,inner sep=0pt,font=\tiny},
  fchk/.style={rectangle,draw,fill=gray!18,minimum size=6mm,inner sep=0pt,font=\tiny},
  Aedge/.style={teal!65!black,line width=0.6pt},
  Bedge/.style={orange!85!black,line width=0.6pt},
  fedge/.style={gray!60,line width=0.6pt},
  ax/.style={gray!55,line width=0.5pt,dashed},
  elab/.style={fill=white,inner sep=1.5pt,font=\small}]
\node[anchor=south west,font=\footnotesize] at (-4.3,3.6) {(a)};
\node[fbit] (a2) at (-2.6, 3) {$[m_A]$};
\node[fchk] (a1) at ( 2.6, 3) {$[n_A]$};
\draw[fedge] (a2) -- (a1) node[midway,fill=white,inner sep=1pt,font=\scriptsize]{$A$};
\node[gray!70,font=\scriptsize] at (0,3.5) {$T_A$};
\node[fbit] (b2) at ( 4.3, 1.9) {$[m_B]$};
\node[fchk] (b1) at ( 4.3,-1.9) {$[n_B]$};
\draw[fedge] (b2) -- (b1) node[midway,fill=white,inner sep=1pt,font=\scriptsize]{$B$};
\node[gray!70,font=\scriptsize,rotate=-90] at (4.8,0) {$T_B$};
\node[data] (VV) at (-2.6, 1.9) {$\mathrm{VV}(j,k)$\\[1pt]$h_1g_x$};
\node[xchk] (X)  at ( 2.6, 1.9) {$X(i,k)$\\[1pt]$g_x$};
\node[zchk] (Z)  at (-2.6,-1.9) {$Z(j,l)$\\[1pt]$h_1g_xh_2$};
\node[data] (CC) at ( 2.6,-1.9) {$\mathrm{CC}(i,l)$\\[1pt]$g_xh_2$};
\draw[ax] (a1) -- (X);   \draw[ax] (a2) -- (VV);
\draw[ax] (b2) -- (X);   \draw[ax] (b1) -- (CC);
\foreach \b in {-10,0,10}{
  \draw[Aedge] (X)  to[bend left=\b] (VV);
  \draw[Aedge] (CC) to[bend left=\b] (Z);
  \draw[Bedge] (X)  to[bend left=\b] (CC);
  \draw[Bedge] (VV) to[bend left=\b] (Z);
}
\node[elab] at (0, 2.5) {\textcolor{teal!45!black}{$[h_1]_{L}\in[a_{ij}]_{L}$}};
\node[elab] at (0,-2.5) {\textcolor{teal!45!black}{$[h_1^{-1}]_{L}\in[a^{*}_{ij}]_{L}$}};
\node[elab,rotate=90] at ( 3.4,0) {\textcolor{orange!70!black}{$[h_2]_{R}\in[b_{kl}]_{R}$}};
\node[elab,rotate=90] at (-3.4,0) {\textcolor{orange!70!black}{$[h_2^{-1}]_{R}\in[b^{*}_{kl}]_{R}$}};
\end{tikzpicture}

\vspace{2mm}

\hspace*{-3mm}%
\begin{tikzpicture}[x=0.62cm,y=0.62cm,>=stealth,
  cell/.style={rounded corners=1.2pt, draw=black!65, minimum width=0.5cm,
               minimum height=0.5cm, inner sep=0pt, font=\scriptsize},
  smid/.style={cell, fill=blue!14},
  cut/.style={cell, dashed, draw=black!45}]
  \definecolor{cA}{RGB}{237,125,49}\definecolor{cB}{RGB}{112,173,71}
  \definecolor{cC}{RGB}{146,111,191}\definecolor{cD}{RGB}{91,155,213}
  \tikzset{k1/.style={cell,fill=cA!55}, k2/.style={cell,fill=cB!55},
           k3/.style={cell,fill=cC!55}, k4/.style={cut,fill=cD!22}}
  \node[anchor=south west,font=\footnotesize] at (-2.6,3.75) {(b)};
  \fill[black!4] (0,0) rectangle (2,3);
  \fill[blue!3]  (2,0) rectangle (6,3);
  \fill[black!4] (6,0) rectangle (8,3);
  \draw[black!25] (2,0)--(2,3) (6,0)--(6,3);
  \draw[<->,black!55] (0.08,3.18)--(1.92,3.18);
  \draw[<->,black!55] (2.08,3.18)--(5.92,3.18);
  \draw[<->,black!55] (6.08,3.18)--(7.92,3.18);
  \node[font=\scriptsize,black!70] at (1,3.55){$E$};
  \node[font=\scriptsize,black!70] at (4,3.55){$M$};
  \node[font=\scriptsize,black!70] at (7,3.55){$L$};
  \foreach \x in {1,...,8} \node[font=\scriptsize,black!55] at (\x-0.5,-0.42){\x};
  \node[font=\scriptsize,anchor=east,align=right] at (-0.2,2.4){$\tau(a_{ij})$\\[-1pt]$=\tau(a_{ij}^{*})$};
  \node[font=\scriptsize,anchor=east] at (-0.2,1.4){$\tau(b_{kl}^{*})$};
  \node[font=\scriptsize,anchor=east] at (-0.2,0.4){$\tau(b_{kl})$};
  \foreach \x/\l in {3/1,4/2,5/3} \node[smid] at (\x-0.5,2.4){$\l$};
  \node[cut,fill=blue!6] at (5.5,2.4){$4$};
  \node[k4] at (0.5,1.4){$4$}; \node[k3] at (1.5,1.4){$3$};
  \node[k2] at (6.5,1.4){$2$}; \node[k1] at (7.5,1.4){$1$};
  \node[k1] at (0.5,0.4){$1$}; \node[k2] at (1.5,0.4){$2$};
  \node[k3] at (6.5,0.4){$3$}; \node[k4] at (7.5,0.4){$4$};
\end{tikzpicture}
\caption{\textbf{LP$(A,B)$ codes.} (a)~Tanner graph $T_{\rm LP}$, with checks $X(i,k)$ (red), $Z(j,l)$ (green) and data qubits $\mathrm{VV}(j,k)$, $\mathrm{CC}(i,l)$ (blue). The second line of each vertex is its group coordinate, its position in the fiber over $T_A~\square~T_B$. Teal bundles run through $[a_{ij}]_L,[a^{*}_{ij}]_L$ and orange bundles through $[b_{kl}]_R,[b^{*}_{kl}]_R$, one edge per group element of the entry. (b)~Sandwich parity-check circuit for $\Delta_A=\Delta_B=4$, with $\mathcal{P}_{\square}$ merging the horizontal edges. Rows give the CNOT layer $\tau$ after edge coloring and entries are the color indices. Dashed cells are present only for even degree: $\Delta_A=3$ shortens $M$ by a layer, whereas $\Delta_B=3$ empties one outer entry on each side without narrowing $E,L$.}
\label{fig: LP code}
\end{figure}

\paragraph*{Disassembling qLDPC codes.--} 
We aim to construct minimal-CNOT-depth parity-check circuits for a qLDPC code with Tanner graph $T_{\mathcal{C}}$. In codes with hundreds of qubits, this problem is generally difficult to solve both analytically and computationally~\cite{zhang2026optimalsyndrome}. 
However, most known good qLDPC codes have substantial structure. Many are assembled from a small set of components through a sequence of operations~\cite{rakovszky2024physicsgoodldpccodes}
\begin{equation}
T_0\xrightarrow{\ \text{Op.\ }1\ }T_1\xrightarrow{\ \text{Op.\ }2\ }\cdots
\xrightarrow{\ \text{Op.\ }N\ }T_{\mathcal C}, \ 
\end{equation}
where each operation is explicitly specified as part of the code definition. Consequently, the edges of $T_{\mathcal C}$ are not independent, but are related by construction~\cite{rakovszky2024physicsgoodldpccodes}. We leverage these relations by grouping related edges and assigning them identical time labels. Specifically, we define an \emph{edge partition} $\mathcal P$: an equivalence relation on the edge set $E$ whose classes $\langle e\rangle_{\mathcal P}$ satisfy
\begin{equation}
\label{eq: equivalent edge time constraint}
e_1,e_2\in\langle e\rangle_{\mathcal P}\ \Rightarrow\ \tau(e_1)=\tau(e_2) \ .
\end{equation}
Eq.~\eqref{eq: incidence edge label constraint} ensures that edges sharing end points belong to distinct classes.

To simplify the parity-check scheduling problem, we choose a sequence of edge partitions $\{\mathcal P_i\}$ that reverses the code-construction operations one at a time:
\begin{equation}
T_0\xleftarrow{\ \mathcal P_1\ }T_1\xleftarrow{\ \mathcal P_{2}\ }\cdots
\xleftarrow{\ \mathcal P_N\ }T_{\mathcal C} \ ,
\end{equation}
where $\mathcal{P}_i$ identifies the edges related by the $i$th operation. 
Each $\mathcal{P}_i$  quotients an edge symmetry imposed during the code assembling, replacing individual edges by edge classes that share a common CNOT layer. To ensure that a solution on the reduced graph remains valid on $T_{\mathcal C}$, Eqs.~\eqref{eq: incidence edge label constraint} and~\eqref{eq: cpsat constraint} must be transformed accordingly under each  $\mathcal P_i$. After applying the complete sequence of partitions, we obtain an edge structure resembling $T_0$ and, crucially, a much smaller scheduling problem.  Its parity-check circuit can  be efficiently found and lifted back to $T_{\mathcal C}$, see Fig.~\ref{fig: growing diagram for qLPDC code}.

\paragraph*{Lifted Product codes.--}
LP codes are constructed in two steps. We take the hypergraph product $\square$ of two protographs $A\in\mathcal M_{n_A\times m_A}$ and $B\in\mathcal M_{n_B\times m_B}$, whose entries $a_{ij},b_{kl}$ lie in the group algebra $\mathbb F_2[G]$, and then perform the group lift~\cite{panteleev2022asymptoticallygoodquantumlocally}
\begin{equation}
\label{eq: construction step in LP}
(T_A,T_B)\xrightarrow{\ \square\ }T_A~\square~T_B\xrightarrow{\ G\text{-lift}\ }T_{\rm LP} \ .
\end{equation}
To ensure the commutation relation between stabilizers, we lift $A$ by the left- and $B$ by the right-regular representation ($\mathcal{L}$ and $\mathcal{R}$, respectively) such that the parity-check matrices of $\mathrm{LP}(A,B)$ read
\begin{subequations}
\begin{align}
H_X&=\begin{bmatrix} \mathcal{L}(A\otimes I_{n_B}) & \mathcal{R}(I_{n_A}\otimes B)\end{bmatrix},\\
H_Z&=\begin{bmatrix}\mathcal{R}(I_{m_A}\otimes B^{*}) &  \mathcal{L}(A^{*}\otimes I_{m_B})\end{bmatrix},
\end{align}
\end{subequations}
where $[A^{*}]_{ij}=a_{ji}^{*}$, with $g\mapsto g^{*}\equiv g^{-1}$ being the antipode map of $\mathbb F_2[G]$. 
Each qubit is indexed by a pair of protograph coordinates and a group coordinate. This code family includes several important special cases: choosing a trivial group $G$ recovers hypergraph product codes, while trivial protographs $A,B\in\mathcal M_{1\times 1}$ yield two-block group algebra codes \cite{lin2023quantumtwoblockgroupalgebra}, including bivariate-bicycle codes \cite{panteleev2022asymptoticallygoodquantumlocally,Panteleev_2022,Bravyi_2024}.

Fig.~\ref{fig: LP code}(a) shows the resulting Tanner graph $T_{\rm LP}$. Two checks $X(i,k,g_x)$ and $Z(j,l,g_z)$ share one VV and one CC data qubit per factorization $g_z=h_1g_xh_2$ with $h_1\in a_{ij}$ and $h_2\in b_{kl}$. Summing over all such factorizations, the properness constraint Eq.~\eqref{eq: cpsat constraint} for this pair reads
\begin{equation}
\label{eq:lp_contam_full main text}
\begin{aligned}
\sum_{\substack{h_1,h_2\\ g_z=h_1g_xh_2}}
\Bigl(&\ind\bigl[\tau([a_{ij}]_{h_1})<\tau([b_{kl}^{*}]_{h_2})\bigr]\\
&+\ind\bigl[\tau([b_{kl}]_{h_2})<\tau([a_{ij}^{*}]_{h_1})\bigr]\Bigr)
\equiv 0\pmod 2 .
\end{aligned}
\end{equation}

The group lift in Eq.~\eqref{eq: construction step in LP} imposes an equivalence relation $\mathcal{P}_{\rm lift}$ identifying edges that carry the same monomial $h_1\in a_{ij}$ ($h_1^{-1}\in a_{ij}^{*}$) or $h_2\in b_{kl}$ ($h_2^{-1}\in b_{kl}^{*}$), e.g.
\begin{equation}
    X(i,k,g_{1})-\mathrm{VV}(j,k,h_1g_{1})\sim
 X(i,k,g_{2})-\mathrm{VV}(j,k,h_1g_{2}) \ .
\end{equation}
Equivalently, $\mathcal{P}_{\rm lift}$ collapses the group coordinate: the resulting edge structure is $T_A~\square~T_B$, with each bundle in Fig.~\ref{fig: LP code}(a) reduced to $W(a_{ij})$ or $W(b_{kl})$ parallel edges, the weight of the corresponding entry. 

Under $\mathcal{P}_{\rm lift}$ we may replace the edges in
Eq.~\eqref{eq:lp_contam_full main text} by their edge classes, which yields the
properness constraint at the level of the edge partition. The set of factorizations
appearing in the sum, however, depends on the group $G$, so a schedule satisfying it need
not remain valid for another member of the family. We therefore impose the stronger
condition that each summand vanish independently,
\begin{equation}
\label{eq: stricter parity LP main text}
\begin{aligned}
&\ind\bigl[\tau([a_{ij}]_{h_1})<\tau([b_{kl}^{*}]_{h_2})\bigr]\\
&+\ind\bigl[\tau([b_{kl}]_{h_2})<\tau([a_{ij}^{*}]_{h_1})\bigr]
\equiv 0\pmod 2 \, ,
\end{aligned}
\end{equation}
which is independent of $G$ and applies to the whole family.

The hypergraph product $\square$ imposes further structure. Both the horizontal edge $h_1\in a_{ij}$ and its conjugate $h_1^{-1}\in a_{ij}^{*}$ originate from the same edge of $T_A$, so we define an equivalence relation $\mathcal{P}_{\square}$ identifying them. The same can be done for vertical edges and $T_B$, however, the two identifications cannot be imposed simultaneously as they force $\tau([a_{ij}]_{h_1})=\tau([a^{*}_{ij}]_{h_1})$ and $\tau([b_{kl}]_{h_2})=\tau([b^{*}_{kl}]_{h_2})$, violating Eq.~\eqref{eq: stricter parity LP main text}. We choose $\mathcal{P}_{\square}$ to identify the horizontal edges with their conjugates; the vertical case is analogous.

If we merge the horizontal edges, there are only two possible solutions to Eq.~\eqref{eq: stricter parity LP main text} 
\begin{equation}
    \begin{cases}
        \tau([b_{kl}]_{h_2}) < \tau([a_{ij}]_{h_1}) = \tau([a_{ij}^{*}]_{h_1}) < \tau([b_{kl}^{*}]_{h_2}), \\
        \tau([b_{kl}^{*}]_{h_2}) < \tau([a_{ij}]_{h_1}) = \tau([a_{ij}^{*}]_{h_1}) < \tau([b_{kl}]_{h_2}).
    \end{cases}
\end{equation}
Therefore, the edge partition $\mathcal{P}_{\square}$ forces $\tau$ into a sandwich structure of three time bands
\begin{equation}
\label{eq:sandwitch-construction}
    \text{[Early band ($E$)] - [Middle band  ($M$)] - [Late band ($L$)]}\ ,
\end{equation}
with the merged horizontal edges in the middle band and the vertical edges in the early and late bands, see Fig.~\ref{fig: LP code}(b). Properness requires that if a vertical edge $h_2\in b_{kl}$ is in $E$, its conjugates $h_2^{-1}\in b_{kl}^{*}$ must be in $L$, and vice-versa.

The CNOT depth follows. Edge coloring $a_{ij},a^{*}_{ij}$ edges of $M$ requires $\Delta_A=\deg(T_A)$ colors, and edge coloring $b_{kl}$ edges requires $\Delta_B=\deg(T_B)$ colors, but the latter must be split between $E$ and $L$. Assigning $\lceil\Delta_B/2\rceil$ colors to $E$ and the rest to $L$ minimizes the total depth (i.e. a balanced split), with  $b_{kl}^{*}$ edges fixed by properness of the circuit. Fig.~\ref{fig: LP code}(b) shows the full schedule. The total depth is $\Delta_A+2\lceil\Delta_B/2\rceil$, equal to the lower bound $\Delta=\Delta_A+\Delta_B$ for $\Delta_B$ even and $\Delta+1$ for $\Delta_B$ odd. Since $\mathcal{P}_{\square}$ can merge either horizontal or  vertical edges, our parity-check circuit attains $\Delta$ whenever at least one of $\Delta_A,\Delta_B$ is even, and $\Delta+1$ when both are odd. In this latter case, extensive numerics for Gross codes~\cite{zhang2026optimalsyndrome,Bravyi_2024} suggest that the minimum $\Delta$ is not achievable and $\Delta+1$ is the true minimum.

In the SM, we show that this sandwich schedule directly extends to balanced product codes and homological products of  quantum and classical codes despite the additional structure each introduces.

\paragraph*{Quantum Tanner codes.--}

QT codes can be understood geometrically as the embedding of classical codewords on a left-right Cayley complex~\cite{leverrier2022quantumtannercodes}. Here, we utilize an algebraic viewpoint~\cite{leverrier2025smallquantumtannercodes} that assembles them by lifting base codes, revealing explicitly the construction operations that our framework relies on.

We choose four classical codes $C_c\subseteq\mathbb F_2^{n_A}$ and $C_c'\subseteq\mathbb F_2^{n_B}$, $c\in\{0,1\}$, with parity- and generator-matrix pairs $(H_c,G_c)$ and $(H_c',G_c')$, and form the base CSS code
\begin{equation}
\label{eq: QT base code}
H_X^{\rm (base)}=\begin{bmatrix}H_0\otimes G_0'\\ H_1\otimes G_1'\end{bmatrix},\quad
H_Z^{\rm (base)}=\begin{bmatrix}G_0\otimes H_1'\\ G_1\otimes H_0'\end{bmatrix}.
\end{equation}
We fix a finite group $G$ and two multisets $\mathcal A=\{a_i\}_{i\in[n_A]}$, $\mathcal B=\{b_j\}_{j\in[n_B]}$ of its elements, with $L_{\mathcal A}$ and $R_{\mathcal B}$ being the corresponding left and right lifts. The parity-check matrices of the QT code read
\begin{subequations}
\begin{align}
H_X&=\begin{bmatrix}H_0\otimes G_0'\otimes I_G\\
(H_1\otimes G_1'\otimes I_G)\,L_{\mathcal A}R_{\mathcal B}\end{bmatrix}\equiv\begin{bmatrix}X_0\\ X_1\end{bmatrix} \ ,\\
H_Z&=\begin{bmatrix}(G_0\otimes H_1'\otimes I_G)\,R_{\mathcal B}\\
(G_1\otimes H_0'\otimes I_G)\,L_{\mathcal A}\end{bmatrix}\equiv\begin{bmatrix}Z_0\\ Z_1\end{bmatrix} \ .
\end{align}
\end{subequations}
The code assembly follows a similar two-step pattern as LP codes:
\begin{equation}
\label{eq: QT code chain}
\substack{C_0,C_1\\ C_0',C_1'}\ \xrightarrow{\ \otimes\ }\
T_{\rm QT}^{\rm (base)} \ \xrightarrow{\ G\text{-lift}\ }\ T_{\rm QT} \ .
\end{equation}

We label checks by their family $X_0,X_1,Z_0,Z_1$, row pair $r\equiv(r_A,r_B)$, and group coordinate $g$.  Data qubits carry base code coordinates $(i,j)$, group coordinate $g$, and sit at the center of stars of four checks, one per family:
\begin{equation}
\label{eq: set of edges QT}
\setlength{\arraycolsep}{3pt}
\begin{array}{c@{}c@{}c@{}c@{}c}
X_0(r,g) & \searrow^{\, I_G} & & \swarrow^{\,L_{\mathcal A}} & Z_1(r,a_ig)\\[2pt]
 & & Q(i,j,g) & & \\[2pt]
Z_0(r,gb_j^{-1}) \quad  & \nearrow_{\,R_{\mathcal B}} & & \nwarrow_{\,L_{\mathcal A}R_{\mathcal B}} & ~\quad X_1(r,a_igb_j^{-1})
\end{array}
\end{equation}
where $r$ ranges over row pairs whose supports contain $i$ and $j$; e.g.\ $X_0(r)$ has check vector $H_0[r_A]\otimes G_0'[r_B]$.

\begin{figure}[t!]\centering
\begin{tikzpicture}[
  x=0.32cm, y=0.62cm,
  Xb/.style={draw=black!55,line width=0.45pt,rounded corners=1.2pt,fill=blue!22},
  Zb/.style={draw=black!55,line width=0.45pt,rounded corners=1.2pt,fill=orange!32},
  Ib/.style={draw=black!45,densely dashed,line width=0.45pt,rounded corners=1.2pt,fill=black!3},
  qt/.style={font=\scriptsize\bfseries,black!70},
  ax/.style={font=\tiny,black!55},
  hd/.style={font=\tiny,black!65}]
\def\PW{10}\def\GAP{2.6}\def\XR{12.6}
\def\band#1#2#3#4#5#6{%
  \draw[#5] ({#1+#3+0.08},{#2-0.37}) rectangle ({#1+#4-0.08},{#2+0.37});
  \node[font=\tiny,inner sep=0pt] at ({#1+(#3+#4)/2},#2) {#6};}

\foreach \ox in {0,\XR}{\foreach \oy in {2.42,-0.38}{
  \draw[black!30,line width=0.4pt,rounded corners=2pt]
    (\ox-0.35,\oy) rectangle (\ox+\PW+0.35,\oy+1.76);}}
\draw[very thick,dashed,black!60] (\PW+\GAP/2,-0.75) -- (\PW+\GAP/2,5.10);
\draw[very thick,dashed,black!60] (-2.5,1.90) -- (\XR+\PW+0.35,1.90);

\node[qt,anchor=north west] at (-0.20,4.12){TL};
\band{0}{2.95}{0}{2}{Zb}{$Z_1$}
\band{0}{2.95}{2}{6}{Xb}{$X_0X_1$}
\band{0}{2.95}{6}{8}{Zb}{$Z_0$}
\band{0}{2.95}{8}{10}{Ib}{\textcolor{black!50}{idle}}
\node[qt,anchor=north west] at (\XR-0.20,4.12){TR};
\band{\XR}{2.95}{0}{1}{Xb}{\scalebox{0.78}{$X_1$}}
\band{\XR}{2.95}{1}{9}{Zb}{$Z_0Z_1$}
\band{\XR}{2.95}{9}{10}{Xb}{\scalebox{0.78}{$X_0$}}
\node[qt,anchor=north west] at (-0.20,1.32){BL};
\band{0}{0.15}{0}{4}{Xb}{$X_0$}
\band{0}{0.15}{4}{6}{Zb}{$Z_0Z_1$}
\band{0}{0.15}{6}{10}{Xb}{$X_1$}
\node[qt,anchor=north west] at (\XR-0.20,1.32){BR};
\band{\XR}{0.15}{0}{2}{Zb}{$Z_0$}
\band{\XR}{0.15}{2}{6}{Xb}{$X_0X_1$}
\band{\XR}{0.15}{6}{8}{Zb}{$Z_1$}
\band{\XR}{0.15}{8}{10}{Ib}{\textcolor{black!50}{idle}}

\foreach \ox in {0,\XR}{
  \draw[black!45,line width=0.4pt] (\ox,-0.62) -- (\ox+\PW,-0.62);
  \foreach \x in {1,...,10}{\draw[black!45,line width=0.35pt] (\ox+\x,-0.62)--(\ox+\x,-0.76);}
  \node[ax,anchor=north] at (\ox+0.5,-0.78){$1$};
  \node[ax,anchor=north] at (\ox+9.5,-0.78){$\Delta$};
  \node[ax,anchor=north] at (\ox+\PW/2,-0.80){CNOT layer};}

\node[hd,anchor=south] at ({(\XR+\PW)/2},5.15) {columns};
\node[hd,anchor=south,align=center] at (\PW/2,4.28)
  {1st half\\[-1pt]\textcolor{black!80}{$X_0\!\prec\!Z_0$,\; $Z_1\!\prec\!X_1$}};
\node[hd,anchor=south,align=center] at (\XR+\PW/2,4.28)
  {2nd half\\[-1pt]\textcolor{black!80}{$Z_0\!\prec\!X_0$,\; $X_1\!\prec\!Z_1$}};
\node[hd,rotate=90,anchor=south] at (-3.05,1.90) {rows};
\node[hd,anchor=east,align=right] at (-0.65,3.05)
  {1st half\\[-1pt]\textcolor{black!80}{$Z_1\!\prec\!X_0$}\\[-1pt]\textcolor{black!80}{$X_1\!\prec\!Z_0$}};
\node[hd,anchor=east,align=right] at (-0.65,0.25)
  {2nd half\\[-1pt]\textcolor{black!80}{$X_0\!\prec\!Z_1$}\\[-1pt]\textcolor{black!80}{$Z_0\!\prec\!X_1$}};
\end{tikzpicture}
\caption{\textbf{Schematic of the canonical orientation for QT codes}, drawn
for $C_i=C_j'=[6,3,3]$ of Ref.~\cite{leverrier2025smallquantumtannercodes} with
$\Delta=10$. Rows and columns of the base surface are halved (dashed lines), and
each half is assigned the orientation listed beside it. In each quadrant the row
and column relations compose into a partial order on the four check families,
so the schedule again has a sandwich structure, with the $X$-families (blue) or
the $Z$-families (orange) in the middle band. The bands show only the order in
which the families act on each data qubit, and their widths are indicative: the
CNOT layers themselves are assigned qubit by qubit by the greedy algorithm .}
\label{fig: quadrant bit orientation}
\end{figure}

Two checks $X_0(r,g_X)$ and $Z_1(r',g_Z)$  overlap on a data qubit $Q(i,j,g)$ iff $g=g_X=a_i^{-1} g_Z$. When the multiset $\mathcal A$ has no repeated elements, this fixes $i$ and leaves $j$ free, such that the two checks overlap along row $i$. Repeated elements admit several $i$, leaving the structure of each  overlap unchanged. The overlap has even size, since $G_0'[r_B]\cdot H_0'[r_B']^{T}=0$, satisfying the commutation relation between the two checks. The same argument applied to each pairing yields
\begin{equation}
\label{eq: QT overlaps}
\begin{aligned}
&(X_0,Z_1),\,(X_1,Z_0)&& \text{overlap along rows },\\
&(X_0,Z_0),\,(X_1,Z_1)&& \text{overlap along columns } .
\end{aligned}
\end{equation}
Each overlap between $X$- and $Z$-checks is thus confined to a union of rows or columns of the $n_A\times n_B$ surface. Indexing an edge uniquely by $(\text{check},r,i,j,g)$, with $g$ being the group coordinate of its check endpoint, and abbreviating $\tau_{X_0}^{r}(i,j,g_X)\equiv\tau(X_0,r,i,j,g_X)$, the properness constraint Eq.~\eqref{eq: cpsat constraint} for the pair $X_0(r,g_X)$, $Z_1(r',g_Z)$ reads
\begin{equation}
\label{eq: QT contamination}
\sum_{\substack{a_i\in\mathcal A\\ g_Z=a_ig_X}} \sum_{j \in S_{\rm row}}
\ind\bigl[\tau_{X_0}^{r}(i,j,g_X)<\tau_{Z_1}^{r'}(i,j,g_Z)\bigr]\equiv 0 \pmod 2 \ ,
\end{equation}
with $S_{\rm row} = \supp (G^{'}_{0}[r_b]) \cap \supp(H_{0}^{'}[r_b^{'}])$, and identically for the other pairs. The outer sum collects the rows/columns on which the pair overlaps, the inner sum the qubits within each of these rows/columns.

The group lift in Eq.~\eqref{eq: QT code chain} imposes the edge partition $\mathcal{P}_{\rm lift}$, which collapses the group coordinate
\begin{equation}
    (\text{check},r,i,j,g_1) \sim (\text{check},r,i,j,g_2) \quad \forall g_1,g_2 \in G \ ,
\end{equation}
making the labels independent of $g$. This does not remove the outer sum in Eq.~\eqref{eq: QT contamination} as it depends on the row index $i$. Therefore, in analogy to Eq.~\eqref{eq: stricter parity LP main text}, we impose the stricter condition that the inner sum vanishes on each row or column independently, yielding
\begin{equation}
\label{eq: QT stricter}
    \sum_{j\in\mathcal S_{\rm row}}
\ind\bigl[\tau_{X_0}^{r}(i,j)<\tau_{Z_1}^{r'}(i,j)\bigr]\equiv 0 \pmod 2 \ ,
\end{equation}
and equivalently for each check pairs. Note that if the multiset $\mathcal{A}$ and $\mathcal{B}$ only have unique elements, then Eq.~\eqref{eq: QT contamination} and Eq.~\eqref{eq: QT stricter} coincide. 

In analogy to LP codes, we construct a schedule $\tau$ that automatically satisfies the properness constraint~\eqref{eq: QT stricter} by declaring an orientation for each row and column and imposing which check type in the overlapping pairs of Eq.~\eqref{eq: QT overlaps} interacts with the data qubits first. For example, for row $i$ we can choose $ X_0 \prec Z_1,~ Z_0 \prec  X_1$. The symbol $\prec$ indicates that for any data qubit in that row, the $X_0$ (and $Z_0$) checks must interact with them before $Z_1$ ($X_1$). Performing edge coloring and labeling of $T_{\rm QT}^{\rm (base)}$ respecting these orientations, we construct a proper parity-check circuit that automatically satisfies Eq.~\eqref{eq: QT stricter}. We remark that the reduced edge-coloring problem depends only on the classical codes $C_c,C_c'$ and not on the group $G$ and the multisets $\mathcal{A},\mathcal{B}$. 

Note that only some orientations are valid, since certain choices for rows $i$ and columns $j$ lead to a contradiction in ordering. A canonical valid choice, which lets all four check families participate in any CNOT layer, is shown in Fig.~\ref{fig: quadrant bit orientation}: we divide the row and column sets into two halves and assign one orientation to the first half and the opposite to the second. The four relations then compose into a partial order in each quadrant, such that for each quadrant the schedule has a sandwich structure similar to~\eqref{eq:sandwitch-construction}.

We apply this strategy to the QT codes of Ref.~\cite{leverrier2025smallquantumtannercodes}. Using the structure of the reduced problem, we design a simple greedy algorithm (Box~\ref{alg: greedy schedule}) to constructs circuit that respect the canonical orientation, and a CP-SAT solver~\cite{cpsatlp} resolves the cases it misses. The resulting circuits reach the lower bound $\Delta$ on all but two of the codes; for those two, CP-SAT certifies that no depth-$\Delta$ circuit respects the canonical orientation, and our circuits have depth $\Delta+1$. Figure~\ref{fig: quadrant bit orientation} sketches the schematic of the circuit. In the SM, we report the full numerical results, including a sweep over permutations of the base codes and a comparison with the left-right circuit of Ref.~\cite{strikis2026syndromeextractioncircuits}.

\paragraph*{Conclusion.--}
We introduced a framework for constructing low-CNOT-depth interleaved parity-check circuits for CSS qLDPC codes that disassembles a code by reading its construction backward: quotienting the symmetry imposed by each operation reduces the scheduling problem to a much smaller instance on the underlying components. For LP codes this yields an analytical circuit within one layer of the lower bound on CNOT depth, and we extend the construction to balanced-product codes and to quantum--classical homological products. For QT codes, a simple greedy algorithm on the reduced problem reaches the lower bound on nearly every instance we test. These results show that the algebraic structure underlying good qLDPC codes can substantially simplify the design of their parity-check circuits.

Our circuits leave considerable freedom in arranging the CNOT gates, so one could enumerate the admissible schedules and retain the one with the largest circuit distance. Two results make this search more tractable. Ref.~\cite{strikis2026syndromeextractioncircuits} associates to each interleaved circuit a non-interleaved counterpart whose circuit distance lower-bounds that of the original and is far cheaper to evaluate, so promising schedules can be identified by screening their counterparts. In a memory experiment, time-reversing the circuit on every second round makes the circuit distance of the full experiment equal to that of a single round~\cite{shaw2026optimisingquantumerrorcorrection}, so only one round needs to be analysed. Together, these reduce the search to a single-round, non-interleaved calculation, which we leave to future work.

\paragraph*{Acknowledgment.--}

We thank Barbara Terhal for valuable feedback on the manuscript, and for pointing us to the circuit-distance result for time-reversed parity-check circuits in Ref.~\cite{shaw2026optimisingquantumerrorcorrection}. During the preparation of this work, the authors used generative artificial intelligence tools. In particular, AI was used to generate TikZ figures and and to assist with numerical code in accordance with the authors’s instructions. All AI-generated outputs were independently verified and reviewed by the authors. The authors retained full responsibility for the scientific content, interpretations, and conclusions of this work.

During the preparation of this manuscript, we became aware of the work of Tan \emph{et al.}~\cite{tan26syndrome}, in which the sandwich schedule for LP codes was obtained independently. We thank the authors for contacting us and refer the reader to their work.

This research was supported by the EU through the H2024 QLSI2 project,  by the Army Research Office under Award Number: W911NF-23-1-0110, and by NCCR Spin (grant number 225153). The views and conclusions contained in this document are those of the authors and should not be interpreted as representing the official policies, either expressed or implied, of the Army Research Office or the U.S. Government. The U.S. Government is authorized to reproduce and distribute reprints for Government purposes notwithstanding any copyright notation herein. M.R.-R. additionally acknowledges support from the Dutch Research Council (NWO) under Award Number Vidi TTW 22204.

\paragraph*{Data availability.-}
The data that support the findings of this study are openly available at \url{https://github.com/Naito27/autext}

\FloatBarrier
\bibliography{report}

\clearpage
\onecolumngrid          
\setcounter{secnumdepth}{3}

\setcounter{equation}{0}
\setcounter{figure}{0}
\setcounter{table}{0}
\setcounter{section}{0}
\renewcommand{\theequation}{S\arabic{equation}}
\renewcommand{\thefigure}{S\arabic{figure}}
\renewcommand{\thetable}{S\arabic{table}}
\renewcommand{\thesection}{S\arabic{section}}
\renewcommand{\thetheorem}{S\arabic{theorem}}   

\clearpage
\begin{center}
  \textbf{\large Supplemental Material for: Disassembling qLDPC codes for
  depth-optimal parity-check circuits}\\[1.2em]
  Minh T.\ P.\ Nguyen, Maximilian Rimbach-Russ, and
  Stefano Bosco\\[0.6em]
\end{center}
\vspace{1em}
\begin{center}
\begin{minipage}{0.88\linewidth}\small
The Supplemental Material contains the analytical derivations and numerical results supporting the main text. Section~\ref{supp: integer programmming formulation} derives the properness constraint and shows that any parity-check circuit satisfying it preserves the stabilizers and logical operators. Section~\ref{supp: lifted product code} details the sandwich schedule for LP codes, and Secs.~\ref{supp: balanced product code} and~\ref{supp: QxC code} extend it to balanced-product codes and to homological products of a quantum and a classical code, the latter with a simple criterion that determines when the resulting depth meets the lower bound up to a single layer. Section~\ref{supp: quantum Tanner code} introduces the canonical orientation and the greedy algorithm for QT codes, and reports their performance on the base codes and on the lifted codes of Ref.~\cite{leverrier2025smallquantumtannercodes}, including a comparison with the left-right circuit of Ref.~\cite{strikis2026syndromeextractioncircuits}.
\end{minipage}
\end{center}
\vspace{1em}

\section{Job scheduling constraint for valid parity check circuit}
\label{supp: integer programmming formulation}

In this section, we show that our parity check circuit $\tau $ is proper, and  preserves the stabilizers and logical operators. Explicitly, we must prove three properties:
\begin{itemize}
  \item \emph{Stabilizer invariance.} The checks $H_X$ and $H_Z$ are invariant under propagation through the parity check circuit
  \begin{equation}
    H_X\xrightarrow{\tau }H_X,\qquad H_Z\xrightarrow{\tau }H_Z.
  \end{equation}

   \item \emph{Logical operators invariance.} The logical operators $L_X$ and $L_Z$ are invariant under propagation through $\tau $
    \begin{equation}
  L_X\xrightarrow{\tau}L_X,\qquad L_Z\xrightarrow{\tau}L_Z.
  \end{equation}

  \item \emph{Properness.} The operator on each $X$-ancilla $v_x$
  (resp.\ $Z$-ancilla $v_z$) propagates through $\tau $ as
  \begin{equation}
  X_{v_x}\xrightarrow{C_{\rm Ext}}X_{v_x}\otimes \prod_{q \in \supp(v_x)  } X_{q} ,\qquad
  Z_{v_z}\xrightarrow{C_{\rm Ext}}Z_{v_z}\otimes\prod_{q \in \supp(v_z)  } Z_q,
  \end{equation}
  with no support on other ancillas.
\end{itemize}
Recall that in the parity check circuit the CNOT gates run from $X$-ancilla (control) to data qubit (target), and from data qubit (control) to $Z$-ancilla (target). Consequently an $X$-type operator propagates through the circuit in the direction
\begin{equation}
\label{eq: X propagation direction}
\text{$X$-ancilla}\;\rightarrow\;\text{data qubit}\;\rightarrow\;\text{$Z$-ancilla},
\end{equation}
while a $Z$-type operator propagates the opposite way,
\begin{equation}
\label{eq: Z propagation direction}
\text{$X$-ancilla}\;\leftarrow\;\text{data qubit}\;\leftarrow\;\text{$Z$-ancilla}.
\end{equation}
 
We first prove stabilizer invariance, giving the argument for $X$-type stabilizers; the $Z$-type case follows by the same reasoning under \eqref{eq: Z propagation direction}. By \eqref{eq: X propagation direction}, the check $v_x$ can propagate only onto $Z$-ancillas. For a given $Z$-ancilla $v_z$, the accumulated $X$-operator is $X_{v_z}$ raised to the parity of $|\operatorname{supp}(v_x)\cap\operatorname{supp}(v_z)|$: each shared data qubit deposits exactly one $X$ on $v_z$ through the single CNOT that joins the shared data qubit to $v_z$, and these mutually commuting contributions are independent of the CNOT layer order. Since every $X$-check commutes with every $Z$-check,
their supports intersect in an even number of data qubits,
\begin{equation}
\bigl|\operatorname{supp}(v_x)\cap\operatorname{supp}(v_z)\bigr|
\equiv 0 \pmod 2,
\end{equation}
so the accumulated operator on $v_z$ is trivial. Hence stabilizers are invariant under the parity check circuit. 

The same argument can also be applied to logical operators $L_X$ and $L_Z$ to prove that they are invariant under the parity check circuit. The key point is that $L_X$ ($L_Z$) overlap $H_Z$ $(H_X$) on an even number of data qubits because of the constraint set by the commutation relation and thus any propagated operator from the logical operators to the ancilla cancels out. 
 
We now show also that properness of the parity check circuit is determined by the relative orderings of the CNOT gates. Consider an $X$-ancilla $v_x$ and a $Z$-ancilla $v_z$ sharing a data qubit $v_q$. We track how an $X$-operator on $v_x$ can reach $v_z$ through $v_q$. The $X$ spreads from $v_x$ onto $v_q$ via the CNOT gate at layer $\tau(v_x,v_q)$, and is then copied onto $v_z$ if and only if $\tau(v_x,v_q)<\tau(v_z,v_q)$, i.e.\ the CNOT on $(v_x,v_q)$ precedes the CNOT on $(v_z,v_q)$; otherwise no such propagation occurs. Summing over the shared qubits, the total operator deposited on $v_z$ is $X$
raised to
\begin{equation}
\label{eq: supp X contamination}
n_{xz}\equiv\!\!\sum_{v_q\,\in\,\operatorname{supp}(v_x)\cap\operatorname{supp}(v_z)}\!\!
\mathbf 1\bigl[\tau(v_x,v_q)<\tau(v_z,v_q)\bigr].
\end{equation}
By the same argument applied along the reversed $Z$-direction
\eqref{eq: Z propagation direction}, a $Z$-operator on $v_z$ deposits onto $v_x$
the operator $Z$ raised to
\begin{equation}
\label{eq: supp Z contamination}
n_{zx}\equiv\!\!\sum_{v_q\,\in\,\operatorname{supp}(v_x)\cap\operatorname{supp}(v_z)}\!\!
\mathbf 1\bigl[\tau(v_x,v_q)>\tau(v_z,v_q)\bigr].
\end{equation}
The two ancillas share an endpoint at each $v_q$, so $\tau(v_x,v_q)\neq \tau(v_z,v_q)$ and the two indicators are complementary; hence $n_{xz}+n_{zx}=|\operatorname{supp}(v_x)\cap\operatorname{supp}(v_z)|$, which is even because every $X$-check commutes with every $Z$-check. Therefore $n_{xz}$ and $n_{zx}$ share the same parity, and thus the pair of checks $(v_x,v_z)$ is proper when
\begin{equation}
\label{eq: cpsat constraint supp}
\sum_{v_q\,\in\,\operatorname{supp}(v_x)\cap\operatorname{supp}(v_z)}
\mathbf 1\bigl[\tau(v_x,v_q)<\tau(v_z,v_q)\bigr]\;\equiv\;0\pmod 2 .
\end{equation}
Consequently, if Eq.~\eqref{eq: cpsat constraint supp} holds for every check pair $(v_x,v_z)$, then the parity check circuit is proper.

\section{Lifted Product code}
\label{supp: lifted product code}
Let $G$ be a finite group and $\mathbb F_2[G]$ its group algebra over the binary
field. An element $a\in\mathbb F_2[G]$ is a formal sum
\begin{equation}
a=\sum_{g\in G}a_g\,g,\qquad a_g\in\mathbb F_2,
\end{equation}
with addition and multiplication defined as.
\begin{equation}
a+b=\sum_{g\in G}(a_g+b_g)\,g,\qquad
ab=\sum_{g\in G}\Bigl(\sum_{hr=g}a_h b_r\Bigr)g.
\end{equation}
The inner sum in multiplication runs over all factorizations $hr=g$ with $h,r\in G$. The group algebra $\mathbb F_2[G]$ carries an \emph{antipode} operation $a\mapsto a^{*}$ defined by $g^{*}=g^{-1}$ and extended linearly through summation and multiplication 
\begin{equation}
a^{*}=\sum_{g\in G}a_g\,g^{-1}
=\sum_{g\in G}a_{g^{-1}}\,g , \qquad (ab)^{*}=b^{*}a^{*}.
\end{equation}
Fix an indexing of the group element $G=\{g_1,\dots,g_\ell\}$ (with $\ell=|G|$), each element $g\in G$ admits two representations, namely the left- and right-regular
\begin{equation}
[g]_{L,\,ij}=
\begin{cases}1,&g_i=g\,g_j,\\0,&\text{else,}\end{cases}
\qquad
[g]_{R,\,ij}=[\rho(g)]_{ij}=
\begin{cases}1,&g_i=g_j\,g,\\0,&\text{else,}\end{cases}
\end{equation}
which are permutation matrices, and extended linearly to $\mathbb F_2[G]$. Both are
faithful, both send the antipode to transpose, $[a^{*}]_L=[a]_L^{T}$ and
$[a^{*}]_R=[a]_R^{T}$, and \emph{the two commute}, $[h]_L[k]_R=[k]_R[h]_L$ for all
$h,k\in G$, since left- and right-translations of $G$ commute.

Given $A\in\mathcal M_{n_A\times m_A}(\mathbb F_2[G])$ and $B\in\mathcal M_{n_B\times m_B}(\mathbb F_2[G])$, the lifted-product code $\mathrm{LP}(A,B)$ is constructed by first taking the hypergraph-product of the two protographs, then lifting $A$ by the left-regular and $B$ by the right-regular representation. Writing $A_L,B_R$ for the lifted matrices, the lifted-product code $\mathrm{LP}(A,B)$ has parity-check matrices
\begin{equation}
H_X=\begin{bmatrix} \mathcal{L}(A\otimes I_{n_B}) & \mathcal{R}(I_{n_A}\otimes B)\end{bmatrix},
\qquad
H_Z=\begin{bmatrix} \mathcal{R}(I_{m_A}\otimes B^{*}) & \mathcal{L}(A^{*}\otimes I_{m_B})\end{bmatrix}.
\end{equation}
The CSS orthogonality $H_XH_Z^{T}=0$ holds identically: since every left-lift $[a_{ij}]_L$ commutes with every right-lift $[b_{kl}]_R$, the mixed-product identity gives \cite{Panteleev_2022}
\begin{equation}
\mathcal{L}(A\otimes I_{n_B})\mathcal{R}(I_{m_A}\otimes B_R)
=\mathcal{R}(I_{n_A}\otimes B_R) \mathcal{L}(A_L\otimes I_{m_B}),
\end{equation}
whence $H_XH_Z^{*} = 0$  for arbitrary protograph $A,B$ due to the binary field. 

Alternatively, we can also view the LP construction as a sequence of operations on the Tanner graph $T_A$ and $T_B$ of the two protographs $A$ and $B$
\begin{equation}
    \label{eq: construction step of LP code}
    T_A, T_B \xrightarrow{\square} T_A ~\square ~T_B \xrightarrow{\text{G-lift}} T_{\rm LP}. 
\end{equation}
Each data qubit and check in the Tanner vertex can be index by a pair of protograph coordinates $(i,j)$ together with a group coordinate $g\in G$ introduced by the lift. The vertex set $V$ of $T_{\rm LP}$ can be naturally split into four sets 
\begin{equation}
\begin{aligned}
\text{VV qubits} &: (j,k,g),\ &&j\in[m_A],\,k\in[n_B];\\
\text{CC qubits} &: (i,l,g),\ &&i\in[n_A],\,l\in[m_B];\\
X\text{-checks}  &: (i,k,g),\ &&i\in[n_A],\,k\in[n_B];\\
Z\text{-checks}  &: (j,l,g),\ &&j\in[m_A],\,l\in[m_B],
\end{aligned}
\end{equation}
see also Fig.~\ref{fig: sub-Tanner-graph LP code supp} in the supplementary material.

\begin{figure}
    \centering
\begin{tikzpicture}[>=latex,
  qubit/.style={circle,draw=black,line width=0.9pt,minimum size=17mm,fill=blue!6,font=\small},
  xchk/.style={rectangle,draw=black,line width=0.9pt,minimum size=13mm,fill=red!12,font=\small},
  zchk/.style={rectangle,draw=black,line width=0.9pt,minimum size=13mm,fill=green!16,font=\small},
  Aedge/.style={teal!65!black,line width=0.7pt},
  Bedge/.style={orange!85!black,line width=0.7pt},
  elab/.style={align=center,inner sep=3pt,fill=white,rounded corners=1.5pt},
  cnt/.style={font=\scriptsize,gray!65,align=center},
]
\node[xchk]  (X) at (0,4.6)   {$X(i,k)$};
\node[qubit] (V) at (7,4.6)   {$\mathrm{VV}(j,k)$};
\node[qubit] (C) at (0,0)     {$\mathrm{CC}(i,l)$};
\node[zchk]  (Z) at (7,0)     {$Z(j,l)$};
\foreach \b in {-15,-5,5,15}{
  \draw[Aedge] (X) to[bend left=\b] (V);
  \draw[Aedge] (C) to[bend left=\b] (Z);
  \draw[Bedge] (X) to[bend left=\b] (C);
  \draw[Bedge] (V) to[bend left=\b] (Z);
}
\node[elab] at (3.5,4.6) {\textcolor{teal!45!black}{$[a_{ij}]_{L}$}\ \ {\footnotesize$\big(W(A)_{ij} \big)$}};
\node[elab] at (3.5,0)   {\textcolor{teal!45!black}{$[a_{ij}^{*}]_{L}$}\ \ {\footnotesize$\big( W(A)_{ij} \big)$}};
\node[elab,rotate=90] at (0,2.3) {\textcolor{orange!70!black}{$[b_{kl}]_{R}$}\ \ {\footnotesize$\big( W(B)_{kl} \big) $}};
\node[elab,rotate=90] at (7,2.3) {\textcolor{orange!70!black}{$[b_{kl}^{*}]_{R}$}\ \ {\footnotesize$\big(W(B)_{kl} \big) $}};
\node[cnt] at (0,5.75)  {$X$-checks\\$n_An_B$ classes};
\node[cnt] at (7,5.75)  {VV qubits\\$m_An_B$ classes};
\node[cnt] at (0,-1.15) {CC qubits\\$n_Am_B$ classes};
\node[cnt] at (7,-1.15) {$Z$-checks\\$m_Am_B$ classes};
\end{tikzpicture}

    \caption{Subgraph of the LP$(A,B)$ Tanner graph spanned by one representative from each register: $X$-checks $X(i,k)$ (red) and $Z$-checks $Z(j,l)$ (green) joined to the data qubits $\mathrm{VV}(j,k)$ and $\mathrm{CC}(i,l)$ (blue). Teal bundles carry the left-lifted entries $[a_{ij}]_L,[a_{ij}^{*}]_L$ of $A$; orange bundles the right-lifted entries $[b_{kl}]_R,[b_{kl}^{*}]_R$ of $B$. Each bundle has multiplicity equal to the corresponding weight-matrix element (in parentheses), one edge per group element in the support of that entry. The registers form a four-cycle due to the hypergraph product construction.}
    \label{fig: sub-Tanner-graph LP code supp}
\end{figure}

From Eq.~\ref{eq: construction step of LP code}, it is straightforward to read-off the natural edge partitions for the LP codes  
\begin{equation}
    \label{eq: natural edge partition of LP code}
    T_A, T_B \xleftarrow{\mathcal{P}_{\square}} T_A ~\square~T_B \xleftarrow{\mathcal{P}_{\rm lift }} T_{\rm LP},
\end{equation}
where $\mathcal{P}_{\rm lift}$ and $\mathcal{P}_{\rm \square}$ invert the group-lifting operation and hypergraph product respectively. The edge partition $\mathcal{P}_{\rm lift}$ is defined by grouping the edges in Fig.~\ref{fig: sub-Tanner-graph LP code supp} by the connecting group element. For instance, the two edges
\begin{equation}
    X(i,k, g_1)\!-\!\mathrm{VV}(j,k,h_1 g_{1})
    \ \sim_{\mathcal{P}_{\rm lift}}\
    X(i,k, g_2)\!-\!\mathrm{VV}(j,k,h_1 g_{2}),
    \qquad h_1 \in \supp(a_{ij}),
\end{equation}
are equivalent because their end points are connected by $h_1$ a term of $a_{ij}$, i.e an $[a_{ij}]_{h_1}$ edge type. In fact, $\mathcal{P}_{\rm lift}$ can also be viewed as a vertex partition by collapsing the the group coordinate of the checks and data qubits. The resulting quotient graph $T_{\rm LP}/\mathcal{P}_{\rm lift} = T_{A}~\square~T_{B}$ is exactly the hypergraph product of the two protographs, whose vertices are index with the protograph indices ($m_An_B$ $\mathrm{VV}$, $n_Am_B$ $\mathrm{CC}$, $n_An_B$ $X$, $m_Am_B$ $Z$). The edges of $T_{A} \square T_{B}$ are the unlifted entries $a_{ij}, b_{kl}$ and their transposes. The number of parallel edges between the vertices is given by the weight of the protograph
\begin{equation}
\begin{aligned}
X(i,k)\!-\!\mathrm{VV}(j,k):&\ \mathrm{wt}(a_{ij}), &\qquad X(i,k)\!-\!\mathrm{CC}(i,l):&\ \mathrm{wt}(b_{kl}),\\
Z(j,l)\!-\!\mathrm{VV}(j,k):&\ \mathrm{wt}(b_{kl}), &\qquad Z(j,l)\!-\!\mathrm{CC}(i,l):&\ \mathrm{wt}(a_{ij}).
\end{aligned}
\label{eq: lp weight edges}
\end{equation}

We now show how the properness parity constraint Eq.~\eqref{eq: cpsat constraint} transform under $\mathcal{P}_{\rm lift}$. On the full Tanner graph, an X-check $(i,k,g_x)$ and a Z-check $(j,l,g_z)$ can share multiple pairs of $(\rm VV,CC)$ data qubit, one pair per factorization $g_z=h_1\, g_x\,h_2$ (with $h_1\in\supp(a_{ij})$ and $h_2\in\supp(b_{kl})$). Explicitly, the $\mathrm{VV}(j,k,h_1 g_x)$ qubit is reached by a $[a_{ij}]_{h_1}$ edge type from the $X$-check and a $[b^{*}_{kl}]_{h_2}$ edge type from the $Z$-check, and the $\mathrm{CC}(i,l)$ qubit is reached by $[b_{kl}]_{h_2}$ and $[a^{*}_{ij}]_{h_1}$ edge types from X- and Z-checks 
\begin{equation}
\label{eq:lp_contam_full}
\sum_{g_x=h_1 g_z h_2}\Bigl(
\mathbf{1}\!\left[\tau([a_{ij}]_{h_1})<\tau([b_{kl}^{*}]_{h_2})\right]
+\mathbf{1}\!\left[\tau([b_{kl}]_{h_2})<\tau([a_{ij}^{*}]_{h_1})\right]\Bigr)\equiv 0 \pmod 2 .
\end{equation}
For a general LP code, all factorization pairs $(h_1 ,h_2)$ and their combinations in Eq.~\eqref{eq:lp_contam_full} are possible. Therefore, we restrict to the parity check circuits satisfy the stricter per-factorization condition that each summand vanishes
\begin{equation}
\label{eq:lp_contam}
\mathbf{1}\!\left[\tau([a_{ij}]_{h_1})<\tau([b_{kl}^{*}]_{h_2})\right]
+\mathbf{1}\!\left[\tau([b_{kl}]_{h_2})<\tau([a_{ij}^{*}]_{h_1})\right]\equiv 0 \pmod 2.
\end{equation}
Assuming that Eq.~\eqref{eq:lp_contam} is true, then Eq.~\eqref{eq:lp_contam_full} is automatically satisfied. Furthermore, we see that Eq.~\eqref{eq:lp_contam} refers to only the edge type and thus is invariant under $\mathcal{P}_{\rm lift}$. 

After tracing out the edge symmetry comes associated with group lifting operation using $\mathcal{P}_{\rm lift}$, we are left with the edge symmetry due to the hypergraph product structure of the code. Specifically, for each edge type $[a_{ij}]_{h_1}$ ($[b_{kl}]_{h_2}$), we have a conjugated counterpart $[a_{ij}^{*}]_{h_1}$ ($[b_{kl}^{*}]_{h_2}$). Therefore, a naive guess of $\mathcal{P}_{\square}$ would be grouping together $[a_{ij}]_{h_1}$ ($[b_{kl}]_{h_2}$) and its conjugate $[a_{ij}^{*}]_{h_1}$ ($[b_{kl}^{*}]_{h_2}$) into a group. However, under this naive grouping, we can verify that Eq.~\eqref{eq:lp_contam} is not satisfied, i.e 
\begin{equation}
    \mathbf{1}\!\left[\tau([a_{ij}]_{h_1})<\tau([b_{kl}^{*}]_{h_2})\right] +\mathbf{1}\!\left[\tau([b_{kl}]_{h_2})<\tau([a_{ij}^{*}]_{h_1})\right]\equiv 1 \pmod 2.
\end{equation}
and  the resulting parity check circuit is not proper. In a valid edge partition $\mathcal{P}_{\square}$, we either group $[a_{ij}]_{h_1}$ and its conjugate $[a_{ij}^{*}]_{h_1}$, or $[b_{kl}]_{h_2}$ and $[b_{kl}^{*}]_{h_2}$, but not both at the same time.

How do we determine whether we merge $[a_{ij}]_{h_1}$ or $[b_{kl}]_{h_2}$ edge types to their conjugates in $\mathcal{P}_{\square}$? A key observation is that no matter which edge types and their conjugate we merge, the resulting contamination parity constraint Eq.~\eqref{eq:lp_contam} naturally forces the schedule $\tau$ to have a sandwich structure with three distinct time bands
\begin{equation}
    \text{[ Early band  ] --- [ Middle band ] --- [ Late band ] },
\end{equation}
where the merged edge types must occupy the middle band. The unmerged edge type occupies the early band and late band, where if an edge type $[b_{kl}]_{h_2}$ occupies the early band, its conjugate $[b_{kl}^{*}]_{h_2}$ must occupy the late band and vice versa. Without loss of generality, let us assume that we merge $[a_{ij}]_{h_1}$ edge types. 

Let us denote the number of colors used in edge coloring of the protograph's Tanner graphs $T_A$ and $T_B$ as $\Delta_A$ and $\Delta_B$ respectively, and thus $T_A~\square~T_B$ requires only $\Delta_A + \Delta_B$ edge colors. As each edge $[a_{ij}]_{h_1}$ or $[b_{kl}]_{h_2}$ and their conjugates can be identified uniquely with an edge in $T_A$ and $T_B$, they can use the same coloring. This also implies that the color $[a_{ij}]_{h_1}$ is identical to its conjugate. $c([a_{ij}]_{h_1})  = c([a_{ij}^{*}]_{h_1})$. As a reminder, edges sharing the same color can be done in the same CNOT layer without any problem (see also Fig.~\ref{fig: sub-Tanner-graph LP code supp}), if we ignore the contamination constraint. 

Performing edge coloring implies that the middle band only requires $\Delta_A$ CNOT layers, as we can trivially stack all $\Delta_A$ color classes of $[a_{ij}]_{h_1}$ and $[a_{ij}^{*}]_{h_1}$. For the early and late band, we need to be more creative. If we use a naive layout where $\Delta_B$ color classes of $[b_{kl}]_{h_2}$ occupy the early band, their conjugates must occupy the late band and the resulting parity check circuit has CNOT depth $\Delta_A + 2 \Delta_B$. 

However, we can divide $\Delta_B$ color classes of $[b_{kl}]_{h_2}$, where $\lceil \Delta_B/2 \rceil$ (here $\lceil \rceil$ is the ceiling function) color classes occupy the early band and the remaining $\Delta_B - \lceil \Delta_B/2 \rceil$ color classes occupy the late band. Due to Eq.~\eqref{eq:lp_contam}, it means that  $\lceil \Delta_B/2 \rceil$ color classes of the conjugate $[b_{kl}^{*}]_{h_2}$ edge type must occupy the late band, and its remaining  $\Delta_B - \lceil \Delta_B/2 \rceil$ color classes must occupy the early band. Therefore, the resulting parity check circuit only uses $\Delta_A + 2 \lceil \Delta_B/2 \rceil$ CNOT layers. If $\Delta_B$ is an even number, our parity check circuit achieves the minimum bound $\Delta_A + \Delta_B$. On the other hand, if $\Delta_B$ is an odd number, our parity check circuit is almost optimal with the CNOT depth of $\Delta_A + \Delta_B +1$. 

The above analysis does not change if we exchange $[a_{ij}^{*}]_{h_1}$ and $[b_{kl}^{*}]_{h_2}$ edge types. Therefore, it answers the question of which edge type, $[a_{ij}^{*}]_{h_1}$ or $[b_{kl}^{*}]_{h_2}$, we merge with their conjugate in the partition $\mathcal{P}_{\square}$. To achieve the minimum CNOT depth parity check circuit, we merge $[b_{kl}^{*}]_{h_2}$ edge types with their conjugate if $\Delta_A$ is even and vice-versa. If both $\Delta_A$ and $\Delta_B$ are odd, both choices lead to identical CNOT depth. Therefore, our sandwich parity check circuit always obtains the minimum CNOT depth circuit $\Delta_A+ \Delta_B$ unless both $\Delta_A$ and $\Delta_B$ are odd. However, for the case that both $\Delta_A$ and $\Delta_B$ are odd, it is unclear whether $\Delta_A + \Delta_B + 1$ is the true optimum CNOT depth if we also need to take properness constraints into account. Recent exhaustive numerical results \cite{zhang2026optimalsyndrome} suggest that it is indeed the true optimum for the Gross code \cite{Bravyi_2024}. 

\FloatBarrier

\section{Balanced Product code}
\label{supp: balanced product code}
\FloatBarrier

\begin{figure}[htb!]
\centering
\begin{tikzpicture}[
  >=Stealth, scale=1, transform shape,
  data/.style={circle,draw,line width=0.8pt,fill=blue!8,minimum size=15mm,inner sep=1pt,
               align=center,font=\scriptsize},
  xchk/.style={rectangle,rounded corners=1pt,draw,line width=0.8pt,fill=red!12,
               minimum width=15mm,minimum height=11mm,inner sep=1pt,align=center,font=\scriptsize},
  zchk/.style={rectangle,rounded corners=1pt,draw,line width=0.8pt,fill=green!14,
               minimum width=15mm,minimum height=11mm,inner sep=1pt,align=center,font=\scriptsize},
  Bas/.style={teal!65!black,line width=0.7pt},
  Fib/.style={orange!85!black,line width=0.7pt},
  elab/.style={font=\scriptsize,inner sep=1.5pt},
  bchip/.style={rounded corners=2pt,draw=teal!55!black,fill=white,line width=0.6pt,
               inner sep=2.5pt,font=\scriptsize\bfseries,text=teal!45!black},
  fchip/.style={rounded corners=2pt,draw=orange!70!black,fill=white,line width=0.6pt,
               inner sep=2.5pt,font=\scriptsize\bfseries,text=orange!60!black}]
 
\begin{scope}
\node[font=\small] at (0,3.95) {(a) one plaquette of $(T_1/H)\otimes_\phi T_2$};
 
\node[data] (VV) at (-3.3, 2.3) {$\mathrm{VV}$\\[1pt]$(\langle v_1\rangle,\, v_2)$};
\node[xchk] (X)  at ( 3.3, 2.3) {$X$\\[1pt]$(\langle c_1\rangle,\, \phi_e\!\cdot\! v_2)$};
\node[zchk] (Z)  at (-3.3,-2.3) {$Z$\\[1pt]$(\langle v_1\rangle,\, c_2)$};
\node[data] (CC) at ( 3.3,-2.3) {$\mathrm{CC}$\\[1pt]$(\langle c_1\rangle,\, \phi_e\!\cdot\! c_2)$};
 
\foreach \b in {-10,0,10}{
  \draw[Bas] (VV) to[bend left=\b] (X);
  \draw[Bas] (Z)  to[bend left=\b] (CC);
  \draw[Fib] (VV) to[bend left=\b] (Z);
  \draw[Fib] (X)  to[bend left=\b] (CC);
}
 
\node[bchip] at (0, 2.3) {split: E\,/\,L};
\node[bchip] at (0,-2.3) {split: E\,/\,L};
\node[fchip] at (-3.3, 0) {merge: M};
\node[fchip] at ( 3.3, 0) {merge: M};
 
\node[elab,text=teal!45!black,anchor=south] at (0, 2.95) {base edge $e\in T_1/H$};
\node[elab,text=teal!45!black,anchor=north] at (0,-2.95) {same base edge $e$, twist $\phi_e$};
\node[elab,text=orange!60!black,anchor=east] at (-4.35, 0.42) {fibre edge};
\node[elab,text=orange!60!black,anchor=east] at (-4.35,-0.02) {$f\in T_2$};
\node[elab,text=orange!60!black,anchor=west] at ( 4.35, 0.42) {partner $\phi_e(f)$:};
\node[elab,text=orange!60!black,anchor=west] at ( 4.35,-0.02) {set by the twist,};
\node[elab,text=orange!60!black,anchor=west] at ( 4.35,-0.46) {not canonical};
 
\draw[<->,teal!60!black,dashed,line width=0.6pt] (0,1.75) -- (0,-1.75);
\node[elab,align=center,text=teal!45!black,fill=white,inner sep=2pt] at (0,0)
     {both copies carry the same $e$:\\ canonical conjugate pair};
\end{scope}
 
\begin{scope}[yshift=-60mm]
\node[font=\small] at (0,1.55) {(b) resulting band structure};
\draw[fill=teal!10,draw=teal!55!black,line width=0.5pt]   (-4.4,0) rectangle (-1.5,0.95);
\draw[fill=orange!12,draw=orange!70!black,line width=0.5pt] (-1.5,0) rectangle ( 1.5,0.95);
\draw[fill=teal!10,draw=teal!55!black,line width=0.5pt]   ( 1.5,0) rectangle ( 4.4,0.95);
\node[font=\scriptsize,text=teal!45!black,align=center]   at (-2.95,0.48) {E\\ base, $\lceil\Delta_1/2\rceil$};
\node[font=\scriptsize,text=orange!60!black,align=center] at ( 0.00,0.48) {M\\ fibre, $\Delta_2$};
\node[font=\scriptsize,text=teal!45!black,align=center]   at ( 2.95,0.48) {L\\ base, $\lceil\Delta_1/2\rceil$};
\draw[->,gray!70,line width=0.5pt] (-4.4,-0.4) -- (4.4,-0.4)
      node[right,font=\scriptsize,text=gray!70] {CNOT layer};
\end{scope}
\end{tikzpicture}
\caption{Tanner graph of the balanced product $T_1\,\square_H\,T_2$ with $H$ free on $T_1$, drawn on one plaquette. Horizontal (teal) edges come from the base graph $T_1/H$; both copies of a plaquette carry the \emph{same} base edge $e$, so base edges have canonical conjugates and can be split between the early and late bands. Vertical (orange) edges come from the fibre $T_2$; the two copies of a plaquette carry $f$ and $\phi_e(f)$, so the partner of a fibre edge is set by the connection and is not canonical. Fibre edges are therefore merged into the middle band.}
\label{fig: BP code}
\end{figure}

The Balanced Product (BP) code \cite{BreuckmannBalancedProduct_2021} is another family of qLDPC codes achieving asymptotically good parameters, and can be viewed as the generalization of the LP code. We consider the balanced product of two classical codes $C_1,C_2$ with Tanner graphs $T_1,T_2$. Let $H$ be a group acting on both graphs by Tanner graph automorphisms, on the right of $T_1$ and on the left of $T_2$. The balanced product $\square_H$ takes the hypergraph product $\square$ and identifies vertices of $T_1~\square~ T_2$ under the anti-diagonal action
\begin{equation}
    \label{eq: BP code symmetry identified}
    (v_1,v_2)\simeq(v_1h^{-1},\,hv_2)\qquad\forall\,h\in H .
\end{equation}
If $H$ acts freely on both factors, the BP code reduces to an LP code \cite{BreuckmannBalancedProduct_2021} and the sandwich circuit applies trivially. We therefore focus on the case where only one factor is free, which we take to be $T_1$. We remark that known good BP instances have at least one free factor \cite{BreuckmannBalancedProduct_2021}.

The edge structure of $T_1\,\square_H\,T_2$ is conveniently described by a \emph{connection} $\phi$ relative to a trivialization $R$ of $T_1$, that is, a choice of representative for each $H$-orbit of vertices. Orient the Tanner edges of $T_1$ from check to bit. For an edge class $\langle e\rangle\equiv\langle c_a\to v_b\rangle$ with representatives $c_a$ and $v_b$ fixed by $R$, the connection $\phi(e)$ is the unique element of $H$ with
\begin{equation}
    c_a\to v_b\cdot\phi(e)\ \in\ \langle c_a\to v_b\rangle .
\end{equation}
Reversing the orientation inverts the connection, $v_b\to c_a\cdot\phi(e)^{-1}$.

We now track how the edges of $T_1\square T_2$ transform under Eq.~\eqref{eq: BP code symmetry identified}. Following the convention of Fig.~\ref{fig: sub-Tanner-graph LP code supp}, we call
\begin{subequations}
\begin{align}
    &\text{VV--}X,\ Z\text{--CC}: && \text{horizontal edges (from } T_1),\\
    &\text{VV--}Z,\ X\text{--CC}: && \text{vertical edges (from } T_2).
\end{align}
\end{subequations}
A horizontal edge transforms as
\begin{equation}
    (\langle c_1\rangle\!\cdot\! h,\ v_2)-(\langle v_1\rangle\!\cdot\!\phi_1 h,\ v_2)
    \ \longrightarrow\ (\langle c_1\rangle,\ h\cdot v_2)-(\langle v_1\rangle,\ \phi_1h\cdot v_2).
\end{equation}
As $h$ ranges over $H$, the second coordinate ranges over the orbit of $v_2$, and $\phi_1$ shifts the fibre coordinate between the two endpoints: the horizontal direction is \emph{twisted}. The same argument gives the twist $\phi^{-1}$ on the $Z$--CC edges. A vertical edge instead transforms as
\begin{equation}
    (\langle v_1\rangle\!\cdot\! h,\ v_2)-(\langle v_1\rangle\!\cdot\! h,\ c_2)
    \ \longrightarrow\ (\langle v_1\rangle,\ h\cdot v_2)-(\langle v_1\rangle,\ h\cdot c_2),
\end{equation}
and, since $H$ acts on $T_2$ by automorphisms, vertical edges are carried to vertical edges within a single fibre. Consequently $T_1\,\square_H\,T_2$ has the fibre bundle structure \cite{MatthewFiberBundle2021}
\begin{equation}
    T_1\,\square_H\,T_2=(T_1/H)\otimes_\phi T_2 ,
\end{equation}
with base graph $T_1/H$, fibre $T_2$, and the connection $\phi$ as the twist, as already noted in Ref.~\cite{BreuckmannBalancedProduct_2021}.

Figure~\ref{fig: BP code} shows that the resulting graph has almost the same shape as the HGP and LP Tanner graphs, but the sandwich schedule cannot be applied naively. Reaching (near-)optimal depth requires a balanced split of an edge type and \emph{its conjugate} between the early and late bands, so the split type must have a canonical conjugate. The two copies of a horizontal edge in a plaquette carry the same base edge $e$, so horizontal edges do; the two copies of a vertical edge carry $f$ and $\phi_e(f)$, so the partner of a vertical edge depends on the base edge and is not canonical. The only consistent assignment therefore places the vertical (fibre) edges in the middle band and splits the horizontal (base) edges between the early and late bands, giving depth
\begin{equation}
    T=\Delta(T_2)+2\Big\lceil \tfrac{\Delta(T_1/H)}{2}\Big\rceil ,
\end{equation}
which meets the lower bound $\Delta=\Delta(T_1/H)+\Delta(T_2)$ when $\Delta(T_1/H)$ is even and exceeds it by one layer otherwise.

\section{Homological product code: quantum $\times$ classical}
\label{supp: QxC code}
Another important family is the homological product of a quantum code $A$ with a classical code $B$ (Q$\times$C), a common routine for balancing code parameters \cite{MatthewFiberBundle2021,Panteleev_2022}; for a thorough review of the chain-complex formulation of homological products we refer the readers to Ref.~\cite{berthusen2025automorphismgadgetshomologicalproduct}. We consider the following question: given a parity-check circuit $\tau_A$ of the quantum code $A$, can we utilize it for a low-depth interleaved parity-check circuit of $A\otimes B$? We show that the answer is yes and quantify the optimality of the construction.

Let $A$ be a CSS code with parity-check matrices $H_X,H_Z$ and let $B$ be a classical code with parity-check matrix $h$. The chain complex of $A\otimes B$ is
\definecolor{Adir}{RGB}{31,95,160}
\definecolor{Bdir}{RGB}{193,90,30}
\definecolor{Bandc}{RGB}{244,244,246}
\begin{equation}
\label{eq: QxC data and check complex}
\begin{tikzpicture}[
    x=1cm,y=1cm,
    >={Stealth[length=2.6mm,width=2.0mm]},
    vert/.style={font=\small,inner sep=2.5pt,fill=white},
    Aar/.style={->,Adir,line width=0.7pt,shorten >=3pt,shorten <=3pt},
    Bar/.style={->,Bdir,line width=0.7pt,shorten >=3pt,shorten <=3pt},
    Alb/.style={font=\footnotesize,Adir,inner sep=2pt},
    Blb/.style={font=\footnotesize,Bdir,inner sep=2pt},
    scale = 0.8
  ]
 
  \foreach \y in {0,2,4,6}{
    \fill[Bandc] (-6.30,\y-0.54) rectangle (9.30,\y+0.54);
  }
 
  \node[vert] (SXB) at ( 0.0,0.0) {$(S_X,B)$};
  \node[vert] (QB)  at (-3.4,2.0) {$(Q,B)$};
  \node[vert] (SXS) at ( 3.4,2.0) {$(S_X,S)$};
  \node[vert] (SZB) at (-3.4,4.0) {$(S_Z,B)$};
  \node[vert] (QS)  at ( 3.4,4.0) {$(Q,S)$};
  \node[vert] (SZS) at ( 0.0,6.0) {$(S_Z,S)$};
 
  \draw[Aar] (SXB) -- node[Alb,below left=-1pt]  {$H_X^{\mathsf T}\!\otimes\mathbf{1}$} (QB);
  \draw[Aar] (QB)  -- node[Alb,left]             {$H_Z\otimes \mathbf{1}$}              (SZB);
  \draw[Aar] (SXS) -- node[Alb,right=3pt]        {$H_X^{\mathsf T}\!\otimes \mathbf{1}$} (QS);
  \draw[Aar] (QS)  -- node[Alb,above right=-1pt] {$H_Z\otimes\mathbf{1}$}              (SZS);
 
  \draw[Bar] (SXB) -- node[Blb,below right=-1pt] {$\mathbf{1} \otimes h$} (SXS);
  \draw[Bar] (SZB) -- node[Blb,above left=-1pt]  {$\mathbf{1} \otimes h$} (SZS);
  \draw[Bar] (QB)  -- node[Blb,fill=white,inner sep=3pt,pos=0.44,sloped,above=0pt] {$\mathbf{1} \otimes h$} (QS);
 
  \node[vert] (TX) at (8.0,0.0) {$\widetilde S_X$};
  \node[vert] (TQ) at (8.0,2.0) {$\widetilde Q$};
  \node[vert] (TZ) at (8.0,4.0) {$\widetilde S_Z$};
  \node[vert] (RZ) at (8.0,6.0) {$\widetilde R_Z$};
 
  \draw[->,line width=0.7pt,shorten >=3pt,shorten <=3pt]
        (TX) -- node[font=\footnotesize,right] {$\widetilde H_X^{\mathsf T}$} (TQ);
  \draw[->,line width=0.7pt,shorten >=3pt,shorten <=3pt]
        (TQ) -- node[font=\footnotesize,right] {$\widetilde H_Z$}             (TZ);
  \draw[->,line width=0.7pt,shorten >=3pt,shorten <=3pt]
        (TZ) -- node[font=\footnotesize,right] {$\widetilde M_Z$}             (RZ);
 
  \draw[gray!55,dashed,line width=0.5pt] (5.9,-0.85) -- (5.9,6.85);
  \node[font=\small\itshape,gray!65] at (0.0,7.05) {double complex $A\otimes B$};
  \node[font=\small\itshape,gray!65] at (8.0,7.05) {total complex};
 
  \foreach \y/\d in {0/0,2/1,4/2,6/3}{
    \node[font=\scriptsize,gray!65,anchor=east] at (-6.25,\y) {$\deg=\d$};
  }
\end{tikzpicture}
\end{equation}
Here $\widetilde S_X,\widetilde S_Z$ and $\widetilde Q$ are the checks and data qubits of the Q$\times$C code, and $\widetilde R_Z$ collects the Z-metachecks, which record the redundancies among the Z-checks and enable tolerance to syndrome-measurement errors. Metachecks are evaluated classically and never measured, so we discard $\widetilde R_Z$ when designing the parity-check circuit. The remaining registers and edge families are
\begin{equation}
\label{eq: QxC schedule}
\begin{tikzpicture}[
    scale=0.8,
    x=1cm,y=1cm,
    baseline=(current bounding box.center),
    >={Stealth[length=2.6mm,width=2.0mm]},
    vert/.style={font=\small,inner sep=2.5pt,fill=white},
    Aar/.style={->,Adir,line width=0.7pt,shorten >=3pt,shorten <=3pt},
    Bar/.style={->,Bdir,line width=0.7pt,shorten >=3pt,shorten <=3pt},
    Alb/.style={font=\footnotesize,Adir,inner sep=2pt},
    Blb/.style={font=\footnotesize,Bdir,inner sep=2pt}
  ]
 
  \foreach \y in {0,2,4}{
    \fill[Bandc] (-6.60,\y-0.54) rectangle (9.80,\y+0.54);
  }
 
  \node[vert] (X)  at ( 0.0,0.0) {$X=(S_X,B)$};
  \node[vert] (Q1) at (-3.6,2.0) {$Q_1=(Q,B)$};
  \node[vert] (Q2) at ( 3.6,2.0) {$Q_2=(S_X,S)$};
  \node[vert] (Z1) at (-3.6,4.0) {$Z_1=(S_Z,B)$};
  \node[vert] (Z2) at ( 3.6,4.0) {$Z_2=(Q,S)$};
 
  \draw[Aar] (X)  -- node[Alb,below left=-1pt] {$d \equiv H_X^{\mathsf T}\!\otimes\mathbf{1}$} (Q1);
  \draw[Aar] (Q1) -- node[Alb,left]            {$u  \equiv H_Z\otimes\mathbf{1}$}              (Z1);
  \draw[Aar] (Q2) -- node[Alb,right=3pt]       {$d' \equiv H_X^{\mathsf T}\!\otimes\mathbf{1}$} (Z2);
 
  \draw[Bar] (X)  -- node[Blb,below right=-1pt] {$b' \equiv \mathbf{1}\otimes h$} (Q2);
  \draw[Bar] (Q1) -- node[Blb,fill=white,inner sep=3pt,pos=0.44,sloped,above=0pt]
                                                {$b \equiv \mathbf{1}\otimes h$} (Z2);
 
  \node[vert] (TX) at (8.2,0.0) {$X$};
  \node[vert] (TQ) at (8.2,2.0) {$Q_1\oplus Q_2$};
  \node[vert] (TZ) at (8.2,4.0) {$Z_1,  Z_2$};
 
  \draw[->,line width=0.7pt,shorten >=3pt,shorten <=3pt]
        (TX) -- node[font=\footnotesize,right] {$\widetilde H_X^{\mathsf T}$} (TQ);
  \draw[->,line width=0.7pt,shorten >=3pt,shorten <=3pt]
        (TQ) -- node[font=\footnotesize,right] {$\widetilde H_Z$}             (TZ);
 
  \draw[gray!55,dashed,line width=0.5pt] (6.1,-0.85) -- (6.1,4.85);
 
  \foreach \y/\d in {0/0,2/1,4/2}{
    \node[font=\scriptsize,gray!65,anchor=east] at (-6.55,\y) {$\deg=\d$};
  }
\end{tikzpicture}
\end{equation}
If we ignore the $Z_1$ checks, the registers $X, Q_1, Q_2, Z_2$ form a hypergraph product code. This suggests reusing the sandwich schedule, with two possible choices
\begin{subequations}
    \begin{equation}
        \centering
\begin{tikzpicture}[x=1mm,y=1mm,font=\footnotesize]
 
\newcommand{\bandrow}[6]{%
  \draw[fill=#5] (#1,#3) rectangle (#2,#4);
  \node at ({(#1+#2)/2},{(#3+#4)/2}) {#6};}
 
\begin{scope}
  \node[font=\small] at (36,23) {(a) Code B in the middle};
  \draw[<->,>=stealth] (13,10) to[bend left=22] node[above,pos=0.5] {balanced-split} (59,10);
  \node at (13,4.5) {early};  \node at (36,4.5) {middle};  \node at (59,4.5) {late};
  \node[anchor=east] at (-1.5,-4.5)  {$Q_1$};
  \node[anchor=east] at (-1.5,-13.5) {$Q_2$};
  \bandrow{0}{26}{0}{-9}{black!7}{$\tau_A$ first half}
  \bandrow{26}{46}{0}{-9}{black!18}{$\mathsf{b}$}
  \bandrow{46}{72}{0}{-9}{black!7}{$\tau_A$ second half}
  \bandrow{0}{26}{-9}{-18}{black!7}{$\mathsf{d}'$}
  \bandrow{26}{46}{-9}{-18}{black!18}{$\mathsf{b}'$}
  \bandrow{46}{72}{-9}{-18}{black!7}{$\mathsf{d}'$}
  \draw[decorate,decoration={brace,amplitude=4pt,mirror}] (0,-19.5) -- (26,-19.5)
        node[midway,below=4pt] {$\lceil |\tau_A|/2\rceil$};
  \draw[decorate,decoration={brace,amplitude=4pt,mirror}] (26,-19.5) -- (46,-19.5)
        node[midway,below=4pt] {$\Delta_B$};
  \draw[decorate,decoration={brace,amplitude=4pt,mirror}] (46,-19.5) -- (72,-19.5)
        node[midway,below=4pt] {$\lceil |\tau_A|/2\rceil$};
\end{scope}
 
\end{tikzpicture}
    \end{equation}
    \begin{equation}
        \begin{tikzpicture}[x=1mm,y=1mm,font=\footnotesize]
            \newcommand{\bandrow}[6]{%
  \draw[fill=#5] (#1,#3) rectangle (#2,#4);
  \node at ({(#1+#2)/2},{(#3+#4)/2}) {#6};}
\begin{scope}[yshift=-56mm]
  \node[font=\small] at (36,23) {(b) Code A in the middle};
  \draw[<->,>=stealth] (8,10) to[bend left=22] node[above,pos=0.5] {balanced-split} (64,10);
  \node at (8,4.5) {early};  \node at (36,4.5) {middle};  \node at (64,4.5) {late};
  \node[anchor=east] at (-1.5,-4.5)  {$Q_1$};
  \node[anchor=east] at (-1.5,-13.5) {$Q_2$};
  \bandrow{0}{16}{0}{-9}{black!18}{$\mathsf{b}$}
  \bandrow{16}{56}{0}{-9}{black!7}{$\tau_A$ \ ($\mathsf{u}$ and $\mathsf{d}$)}
  \bandrow{56}{72}{0}{-9}{black!18}{$\mathsf{b}$}
  \bandrow{0}{16}{-9}{-18}{black!18}{$\mathsf{b}'$}
  \bandrow{16}{56}{-9}{-18}{black!7}{$\mathsf{d}'$}
  \bandrow{56}{72}{-9}{-18}{black!18}{$\mathsf{b}'$}
  \draw[decorate,decoration={brace,amplitude=4pt,mirror}] (0,-19.5) -- (16,-19.5)
        node[midway,below=4pt] {$\lceil \Delta_B/2\rceil$};
  \draw[decorate,decoration={brace,amplitude=4pt,mirror}] (16,-19.5) -- (56,-19.5)
        node[midway,below=4pt] {$|\tau_A|$};
  \draw[decorate,decoration={brace,amplitude=4pt,mirror}] (56,-19.5) -- (72,-19.5)
        node[midway,below=4pt] {$\lceil \Delta_B/2\rceil$};
\end{scope}
        \end{tikzpicture}
    \end{equation}
\end{subequations}
Both variants replace the CNOT gates on the $\mathsf d$-edges by the circuit $\tau_A$, replayed once per bit of $B$. Properness then splits into two families of check pairs. A pair $(X,Z_1)$ overlaps only on $Q_1$ qubits and reproduces, for each bit of $B$, the contamination condition of $A$ itself; replaying $\tau_A$ satisfies it without any knowledge of the algebraic structure of $A$. A pair $(X,Z_2)$ overlaps on one $Q_1$ and one $Q_2$ qubit and is handled by the sandwich bands exactly as for LP codes. The bands are disjoint and each band is properly colored, so no edge collisions occur. Variant (a) has depth $2\lceil|\tau_A|/2\rceil+\Delta_B$ and variant (b) has $|\tau_A|+2\lceil\Delta_B/2\rceil$; taking the better of the two,
\begin{equation}
    T \;=\; |\tau_A| + \Delta_B + \big[\,|\tau_A|\text{ odd}\wedge\Delta_B\text{ odd}\,\big].
\end{equation}
In the following we set $|\tau_A|=\Delta_A$, the minimum depth for the input codes of interest.

To analyze the optimality of the schedule, we count vertex degrees in the chain complex. We abbreviate the maximal degrees
\begin{equation}
    x\equiv\deg(S_X),\quad q\equiv\deg(Q),\quad z\equiv\deg(S_Z),\quad
    \beta\equiv\deg(B),\quad \sigma\equiv\deg(S),
\end{equation}
and write $q_X$ for the maximal degree of $Q$ restricted to the subgraph $Q$--$S_X$. Reading the complex row by row and discarding the edges into $\widetilde R_Z$,
\begin{equation}
\label{eq: QxC general degree counting}
    \deg(X)=x+\beta,\quad \deg(Z_1)=z,\quad \deg(Z_2)=q_X+\sigma,\quad
    \deg(Q_1)=q+\beta,\quad \deg(Q_2)=x+\sigma .
\end{equation}
Every parity-check circuit has depth at least as large as the largest vertex degree of the Tanner graph,
\begin{equation}
    \Delta=\max\big[\,x+\beta,\; q+\beta,\; x+\sigma,\; z,\; q_X+\sigma\,\big],
\end{equation}
while the input codes fix $\Delta_A=\max[x,q,z]$ and $\Delta_B=\max[\beta,\sigma]$. Comparing $T$ with $\Delta$ yields the following regimes:
\begin{itemize}
    \item If $z\le\max[q,x]$ and $\sigma\le\beta$, then $T=\max[q,x]+\beta\le\Delta$ and the schedule is optimal up to the parity layer: the left column of the complex dominates.
    \item If $z\le\max[q,x]$ and $\sigma>\beta$, the schedule remains optimal when $q\le x$, since then $T=x+\sigma=\deg(Q_2)$. When $q>x$ the gap is bounded by
    \begin{equation}
        T-\Delta \;\le\; \min\big[\,q-q_X,\;\sigma-\beta\,\big],
    \end{equation}
    with equality $T-\Delta=q-q_X$ when $\Delta=\deg(Z_2)$; the gap closes whenever $q_X=q$, which depends on the construction of $A$.
    \item The analysis for $z>\max[q,x]$ is analogous. The worst case is $\Delta=\deg(Z_1)=z$, where the gap reaches $\Delta_B$: the $Z_1$ checks lose their $B$-direction edges to the discarded metacheck, and the schedule idles accordingly.
\end{itemize}

The case of our interest is $A$ a HGP code, built from classical codes with bit and check registers $(B_i,S_i)$, $i\in\{1,2\}$, and $B$ a classical code with registers $(B_3,S_3)$. We orient the product so that $S_X=(S_1,B_2)$, $S_Z=(B_1,S_2)$ and $Q=(B_1,B_2)\oplus(S_1,S_2)$, and abbreviate $\beta_i\equiv\deg(B_i)$, $\sigma_i\equiv\deg(S_i)$, $\Delta_i=\max[\beta_i,\sigma_i]$. Then
\begin{equation}
\label{eq: HGPxclas degree supp}
    x=\sigma_1+\beta_2,\qquad z=\beta_1+\sigma_2,\qquad
    q=\max[\,\sigma_1+\sigma_2,\;\beta_1+\beta_2\,],\qquad
    q_X=\max[\,\beta_1,\;\sigma_2\,].
\end{equation}
Maximizing over the four sign combinations gives $\Delta_A=\Delta_1+\Delta_2$, hence
\begin{equation}
    T=\Delta_1+\Delta_2+\Delta_3=\sum_i\max(\beta_i,\sigma_i)
\end{equation}
up to the parity layer, and optimality reduces to the question of when $\Delta=\sum_i\Delta_i$. Substituting Eq.~\eqref{eq: HGPxclas degree supp} into Eq.~\eqref{eq: QxC general degree counting} results in 
\begin{equation}
\begin{aligned}
    \deg(X)&=\sigma_1+\beta_2+\beta_3, &\quad \deg(Z_1)&=\beta_1+\sigma_2, &\quad \deg(Z_2)&=\max[\beta_1,\sigma_2]+\sigma_3,\\
    \deg(Q_1)&=\max[\sigma_1+\sigma_2,\,\beta_1+\beta_2]+\beta_3, &\quad
    \deg(Q_2)&=\sigma_1+\beta_2+\sigma_3. &&
\end{aligned}
\end{equation}
Only $X$, $Q_1$ and $Q_2$ carry one weight from each of the three factors; writing which side of factor $i$ contributes, their side patterns are
\begin{equation}
    (S,B,B),\qquad (S,S,B),\qquad (B,B,B),\qquad (S,B,S).
\end{equation}
These are exactly the patterns that differ in at least two positions from the pattern $(B,S,S)$; the remaining patterns $(B,S,B)$, $(B,B,S)$, $(S,S,S)$ are realized by $Z_1$ and $Z_2$, which lose one factor's weight to the truncation of the discarded metacheck $(S_Z,S)$. Consequently $\Delta=\sum_i\Delta_i$, and the sandwich lift is optimal up to the parity layer, if and only if the heavy sides of the three factors match one of the four patterns above or equivalently, iff at most one of
\begin{equation}
    \beta_1>\sigma_1,\qquad \sigma_2>\beta_2,\qquad \sigma_3>\beta_3
\end{equation}
holds. We remark that our analysis is based on the assumption that the extended logical axis in the QxC code is the Z-axis (i.e. it has metachecks). However, the chain complex Eq.~\eqref{eq: QxC data and check complex} can be viewed in the opposite direction in which the X-axis carries the metachecks. Our analysis can still be applied with minor modifications. 

\section{Quantum Tanner code}
\label{supp: quantum Tanner code}

In this section, we consider the construction of the \textit{quadripartite} Quantum Tanner code as the lifting of a base CSS code \cite{leverrier2025smallquantumtannercodes, leverrier2022quantumtannercodes}. Let $C_0,C_1 \subseteq \mathbb{F}_2^{n_A}$ and $C_{0}^{'}, C_1^{'} \subseteq \mathbb{F}_{2}^{n_B}$ be unprimed and primed binary classical codes with length $n_A$ and $n_B$ respectively. We denote $G_i,H_i$ ($G_i^{'},H_i^{'}$) the generator and parity check matrices of the unprimed (primed) classical codes. We remark that the classical codes (e.g. $C_0^{'}=C_1^{'}$) can be identical, but their parity matrices $(H^{'}_{0},H^{'}_{1})$ and generators $(G^{'}_{0},G^{'}_{1})$ must differ. The parity checks matrices of the base quantum code are given by the \textit{tensor product} of the classical codes 
\begin{equation}
\label{eq: QT base code supp}
H_X^{\rm (base)}=\begin{bmatrix}H_0\otimes G_0'\\ H_1\otimes G_1'\end{bmatrix} \equiv \begin{bmatrix}
    X_0^{\text{(base)}} \\ X_1^{\text{(base)}}
\end{bmatrix} ,~ \quad 
H_Z^{\rm (base)}=\begin{bmatrix}G_0\otimes H_1'\\ G_1\otimes H_0'\end{bmatrix} \equiv \begin{bmatrix}
    Z_0^{\text{(base)}} \\ Z_1^{\text{(base)}}
\end{bmatrix}.
\end{equation}
For a given base check family (e.g. $X_0$), we can index its members by a pair $(r_a,r_b)$, where $r_a$ and $r_b$ are the indexes of the corresponding classical checks. Similarly, we can index a base qubit by a pair $(i,j)$ corresponding to the index of the classical bit. We will refer to $(i,j)$ as a square in a $n_A \times n_B$ surface. 

To lift the base code to a Quantum Tanner code, we choose a group $G$ along with two multisets $A = \{ a_{i} \}_{i \in n_A}$ and $B = \{ b_{j} \}_{j \in n_B}$. The multisets can contain repeated elements, which we will discuss more in detail later. Each base check with index $(r_a,r_B)$ and data qubit with index $(i,j)$ is now replaced by $|G|$ copies, which can be distinguish by appending an additional group index to the base index. We denote the left-regular action $\lambda_a \in \mathbb{F}_{2}^{|G|\times |G|}$ for $a \in A$ and the right-regular action  $\rho_b \in \mathbb{F}_{2}^{|G|\times |G|}$ for $b \in B$ where their actions are given by
\begin{equation}
    [\lambda_a]_{g,h} = \delta_{h, a^{-1} g}, \qquad  [\rho_{b}]_{h,g} = \delta_{h,g b}. 
\end{equation}
From the left- and right-regular action, we can define the block-diagonal permutations $L_A$ and $R_B$ describing the lift action of the multisets $A$ and $B$ 
\begin{equation}
    L_A = \bigoplus_{i=1}^{n_A} \bigoplus_{j=1}^{n_b} \lambda_{a_i}, \quad  R_B = \bigoplus_{i=1}^{n_A} \bigoplus_{j=1}^{n_b} \rho_{b_j}. 
\end{equation}
The parity check matrices of the Quantum Tanner code are then given by 
\begin{align}
&H_X=\begin{bmatrix}H_0\otimes G_0'\otimes I_G\\
(H_1\otimes G_1'\otimes I_G)\,L_{A}R_{B}\end{bmatrix} \equiv \begin{bmatrix}
    X_0 \\ X_1
\end{bmatrix}, \qquad  H_Z=\begin{bmatrix}(G_0\otimes H_1'\otimes I_G)\,R_{B}\\
(G_1\otimes H_0'\otimes I_G)\,L_{A}\end{bmatrix} \equiv \begin{bmatrix}
    Z_0 \\ Z_1
\end{bmatrix}.
\end{align}
The actions of $L_A$ and $R_B$ on the edges of Tanner graph can be read-off directly from the parity check matrices 
\begin{equation}
\begin{aligned}
& X_0(r_a,r_b,g_X)\,\xrightarrow{\,I_{G}}\, Q(i,j,\,g_X),
&&Z_1(r_a,r_b,g_Z)\,\xrightarrow{\,L_{\mathcal A}\,}\, Q(i,j,\,a_i^{-1}g_Z),\\
&Z_0(r_a,r_b,g_Z)\,\xrightarrow{\,R_{\mathcal B}\,}\, Q(i,j,\,g_Z\,b_j),
&&X_1(r_a,r_b,g_X)\,\xrightarrow{\,L_{\mathcal A}R_{\mathcal B}\,}\, Q(i,j,\,a_i^{-1}g_X\,b_j).
\end{aligned}
\end{equation}
Therefore, we can view the construction of the Quantum Tanner code as the chain 
\begin{equation}
\label{eq: QT code chain supp}
\substack{C_0,C_1\\ C_0',C_1'}\ \xrightarrow{\ \otimes\ }\
T_{\rm QT}^{\rm (base)} \ \xrightarrow{\ G\text{-lift}\ }\ T_{\rm QT}. 
\end{equation}

An important property of the Quantum Tanner code is that a X-check and a Z-check can only overlap along either a row $i$ or a column $j$. To be specific, let us examine the checks $X_0(r_a,r_b,g_x)$ and $Z_1(r_a^{'},r_b^{'},g_z)$. They can only overlap at a qubit $Q(i,j,g)$, with 
\begin{equation}
    i \in \supp(H_0[r_A]) \cap \supp(G_1[r_a^{'}]) \quad   j \in \supp(G_0^{'}[r_b]) \cap \supp(H_0^{'}[r_b^{'}]) ),
\end{equation}
iff $g= g_x = a_{i}^{-1} g_z$, and thus fix the row coordinate $i$ (assuming the element $a_i$ has multiplicity one in the multiset A) but leave the column coordinate $j$ free. Along the row $i$, the X-check $X_0(r_a,r_b,g_x)$ has support on the codeword $G_{0}^{'}[r_b]$ whereas the Z-check  $Z_1(r_a^{'},r_b^{'},g_z)$ has support on the codeword $H_{0}^{'}[r_b^{'}]$. Therefore, the commutation relations are automatically satisfied by construction. Let us denote the edges in the Tanner graph as 
\begin{equation}
    (\text{check},r_A,r_B,g,i,j) \quad \text{check} \in \{X_0,X_1,Z_0,Z_1\},
\end{equation}
which details which check family that edge belongs to and its ending data qubit. The properness parity constraint between $X_0(r_a,r_b,g_x)$ and $Z_1(r_a^{'},r_b^{'}, g_z  )$ is 
\begin{equation}
    \label{eq: QT contamination full sup}
    \sum_{j \in \supp(G_0^{'}[r_b]) \cap \supp(H_0^{'}[r_b^{'}])} \mathbf{1} \Big[ \tau\big[(X_0,r_a,r_b,g_x,i,j) \big] < \tau \big[ (Z_1,r_a^{'},r_b^{'}, g_z,i,j)  \big] \big) \Big] = 0 \mod 2 \quad. 
\end{equation}
If $a_i$ has multiplicity greater than one, it means that  $X_0(r_a,r_b,g_x)$ and $Z_1(r_a^{'},r_b^{'},g_z)$ can overlap on more than one row $i$, and the parity constraint Eq.~\eqref{eq: QT contamination full sup} will include an outer sum over the overlapping rows. However, similar to the case of LP code, we will restrict to solutions in which the inner sum (i.e. Eq.~\eqref{eq: QT contamination full sup}) vanish identically. The analysis for other pairs of X-check and Z-check follows accordingly. To summarize, the overlapping direction of check pairs is
\begin{equation}
\label{eq: orientation in QT}
\begin{aligned}
&(X_0,Z_1),\,(X_1,Z_0):&& \text{overlapping along rows}, \\
&(X_0,Z_0),\,(X_1,Z_1):&& \text{overlapping along columns} .
\end{aligned}
\end{equation}

As the first step, we invert the group lift in Eq.~\eqref{eq: QT code chain supp} by collapsing the group coordinate, resulting in the edge partition $\mathcal{P}_{\rm lift}$. Consequently, the edge classes after applying $\mathcal{P}_{\rm lift}$ are 
\begin{equation}
    (\text{check},r_A,r_B,g,i,j) \xrightarrow{\mathcal{P}_{\rm lift}} (\text{check},r_A,r_B,i,j).
\end{equation}
After inverting the group lift, we are left with the base code Eq.~\eqref{eq: QT base code supp}. However, due to how the classical codes are glued together in the tensor product, we did not find a natural way to partition the edges of $T^{(\rm base)}_{\rm QT}$ into the four classical codes $C_i,C_i^{'}$. In principle, we can stop at this step and solve the integer programming problem (e.g. using CP-SAT) to find a depth-optimal interleaved parity check circuit. However, the number of constraints Eq.~\eqref{eq: QT contamination full sup} can remain relatively large, and the solver might struggle as we increase the size of the classical codes $C_i,C_i^{'}$. Therefore, in the following, we will derive a construction such that the circuit is always proper without needing to know the overlapping patterns between the checks, and an efficient heuristic algorithm to find the circuit.  

Specifically, due to the tensor product structure of the codes, we can choose an \textit{orientation} for each rows and columns in the $n_A \times n_B$ surface. For instance, we can choose the following orientation for a row $i$
\begin{equation}
    \label{eq: example orientation supp}
    X_0 \prec Z_1,~ Z_0 \prec  X_1.
\end{equation}
Eq.~\eqref{eq: example orientation supp} means that for any given data qubit in row $i$, the checks $X_0~(Z_0)$ must interact with it first before $Z_1~(X_1)$. Consequently, a schedule following the orientation Eq.~\eqref{eq: example orientation supp}, it also satisfies the parity constraint Eq.~\eqref{eq: QT contamination full sup} without needing to know the overlapping pattern between the classical codes. However, not all choices of orientation are chronologically valid. As an example, let us choose the following orientation for the row $i$ and the column $j$
\begin{subequations}
    \label{eq: QT invalid orientation}
    \begin{align}
        & \text{ row i:} \quad  &X_0 \prec Z_1,~ X_1 \prec  Z_0 \\
        & \text{ column j:} & Z_0 \prec X_0, ~ Z_1 \prec X_1 
    \end{align}
\end{subequations}
then at the data qubit coordinate $(i,j)$, it leads to the requirement $X_0 \prec Z_1 \prec X_1 \prec Z_0 \prec X_0$ which is a contradiction. 

Let us draw out geometrically how the choices can couple the check families at a data qubit
\begin{equation}
  \label{eq: QT orientation cycle}
  \begin{tikzpicture}[baseline=(current bounding box.center),
      fam/.style={rounded corners=2pt, draw=black!55, minimum width=11mm,
                  minimum height=7mm, inner sep=2pt, font=\small},
      xfam/.style={fam, fill=red!12},
      zfam/.style={fam, fill=green!14},
      harp/.style={-{Stealth[harpoon, length=2.6mm]}, very thick},
      row/.style={harp, teal!70!black},
      col/.style={harp, orange!85!black},
      rlbl/.style={font=\small, teal!70!black, inner sep=2pt},
      clbl/.style={font=\small, orange!85!black, inner sep=2pt}]
 
    \node[xfam] (X0) at (0,2.6)   {$X_0$};
    \node[zfam] (Z1) at (5.2,2.6) {$Z_1$};
    \node[xfam] (X1) at (5.2,0)   {$X_1$};
    \node[zfam] (Z0) at (0,0)     {$Z_0$};
 
    \draw[dashed, black!30] (X0) -- (X1);
    \draw[dashed, black!30] (Z0) -- (Z1);
    \node[font=\scriptsize, black!45, fill=white, inner sep=1.5pt]
      at (2.6,1.3) {no ordering};
 
    \draw[row, transform canvas={yshift= 2.2pt}] (X0) -- (Z1);
    \draw[row, transform canvas={yshift=-2.2pt}] (Z1) -- (X0);
    \node[rlbl, anchor=south] at (2.6, 2.80) {$\mathrm{rb}_0 = 1$};
    \node[rlbl, anchor=north] at (2.6, 2.40) {$\mathrm{rb}_0 = 0$};
 
    \draw[row, transform canvas={yshift=-2.2pt}] (X1) -- (Z0);
    \draw[row, transform canvas={yshift= 2.2pt}] (Z0) -- (X1);
    \node[rlbl, anchor=north] at (2.6,-0.20) {$\mathrm{rb}_1 = 1$};
    \node[rlbl, anchor=south] at (2.6, 0.20) {$\mathrm{rb}_1 = 0$};
 
    \draw[col, transform canvas={xshift= 2.2pt}] (Z1) -- (X1);
    \draw[col, transform canvas={xshift=-2.2pt}] (X1) -- (Z1);
    \node[clbl, rotate=90, anchor=north] at (5.42,1.3) {$\mathrm{cb}_1 = 1$};
    \node[clbl, rotate=90, anchor=south] at (4.98,1.3) {$\mathrm{cb}_1 = 0$};
 
    \draw[col, transform canvas={xshift=-2.2pt}] (Z0) -- (X0);
    \draw[col, transform canvas={xshift= 2.2pt}] (X0) -- (Z0);
    \node[clbl, rotate=90, anchor=south] at (-0.22,1.3) {$\mathrm{cb}_0 = 1$};
    \node[clbl, rotate=90, anchor=north] at ( 0.22,1.3) {$\mathrm{cb}_0 = 0$};
  \end{tikzpicture}
\end{equation}
where the binary variables $\{ \mathrm{rb}_i,\mathrm{cb}_j \} $ denote the orientation
\begin{subequations}
  \label{eq: QT orientation bits}
  \begin{align}
    \text{row } i: &\quad
      \mathrm{rb}_0 = \begin{cases} 1, & X_0 \prec Z_1 \\ 0, & Z_1 \prec X_0 \end{cases}
      \qquad
      \mathrm{rb}_1 = \begin{cases} 1, & X_1 \prec Z_0 \\ 0, & Z_0 \prec X_1 \end{cases} \\[2pt]
    \text{column } j: &\quad
      \mathrm{cb_1} = \begin{cases} 1, & Z_1 \prec X_1 \\ 0, & X_1 \prec Z_1 \end{cases}
      \qquad
      \mathrm{cb_0} = \begin{cases} 1, & Z_0 \prec X_0 \\ 0, & X_0 \prec Z_0 \end{cases}
  \end{align}
\end{subequations}
Thus, there are in general $2^{4} = 16$ combinations of orientation choices at a data qubit. Out of these combinations, we see that when all binary variables $\{ \mathrm{rb}_i,\mathrm{cb}_j \} $ are identical (i.e. all one or zero), Eq.~\eqref{eq: QT orientation cycle} forms a directed cycle and there will be inconsistency in the chronological order. The previous example Eq.~\eqref{eq: QT invalid orientation} is a showcase. Other combinations of $\{ \mathrm{rb}_i,\mathrm{cb}_j \}$ are all chronologically valid. 

We now present a canonical choice of row and column orientations. We first divide the row set into two halves and similarly for the column set. We then choose the following orientation for each halves
\begin{equation}
\label{eq: QT canonical orientation supp}
\newcommand{\qposet}[7]{%
  \begin{scope}[shift={(#1,#2)}]
    \draw[black!35, line width=0.4pt, rounded corners=2pt]
      (-0.30,-0.72) rectangle (3.45,0.80);
    \node[qlbl] at (0.02,0.58) {#3};
    \node[fnode] at (0.15, 0.00) {$#4$};
    \node[rel]   at (0.85, 0.00) {$\prec$};
    \node[fnode] at (1.55, 0.34) {$#5$};
    \node[fnode] at (1.55,-0.34) {$#6$};
    \node[rel]   at (2.25, 0.00) {$\prec$};
    \node[fnode] at (2.95, 0.00) {$#7$};
  \end{scope}}
    \begin{tikzpicture}[baseline=(current bounding box.center),
  fnode/.style={draw=black!55, rounded corners=1.5pt, inner sep=1.6pt,
                minimum width=6.5mm, minimum height=4.2mm, font=\small},
  rel/.style={font=\small, black!75},
  qlbl/.style={font=\scriptsize\bfseries, black!60},
  hd/.style={font=\scriptsize, black!65, align=center}]
 
  \qposet{0.0}{2.6}{TL}{Z_1}{X_0}{X_1}{Z_0}
  \qposet{5.0}{2.6}{TR}{X_1}{Z_0}{Z_1}{X_0}
  \qposet{0.0}{0.0}{BL}{X_0}{Z_0}{Z_1}{X_1}
  \qposet{5.0}{0.0}{BR}{Z_0}{X_0}{X_1}{Z_1}
 
  \draw[dashed, black!60, line width=0.7pt] (4.08,-1.10) -- (4.08,4.00);
  \draw[dashed, black!60, line width=0.7pt] (-3.30,1.30) -- (8.45,1.30);
 
  \node[hd, anchor=south] at (4.08,4.20) {columns};
  \node[hd, anchor=south] at (1.58,3.55)
    {1st half\\ $X_0 \prec Z_0$,\; $Z_1 \prec X_1$};
  \node[hd, anchor=south] at (6.58,3.55)
    {2nd half\\ $Z_0 \prec X_0$,\; $X_1 \prec Z_1$};
 
  \node[hd, rotate=90, anchor=south] at (-3.55,1.30) {rows};
  \node[hd, anchor=east] at (-0.45,2.60)
    {1st half\\ $Z_1 \prec X_0$\\ $X_1 \prec Z_0$};
  \node[hd, anchor=east] at (-0.45,0.00)
    {2nd half\\ $X_0 \prec Z_1$\\ $Z_0 \prec X_1$};
\end{tikzpicture}
\end{equation}
where in each quadrant, the partial orders on each data qubit again exhibit the three-band structure. The canonical orientation is chosen such that at any given time band (E,M,L), all the check families can participate in the schedule, and the middle band has flexibility (e.g. the $X_0$ check and $X_1$ checks can intertwine with each other and there are two quadrant to choose from). As a result, if we assign a time label for the base code $T_{\rm QT}^{\rm (base)}$ respecting Eq.~\eqref{eq: QT canonical orientation supp}, it is also proper.

\begin{algbox*}{Greedy interleaved schedule for QT code}
\label{alg: greedy schedule}
\begin{algorithmic}[1]
\Require base Tanner graph $T^{(\mathrm{base})}_{\mathrm{QT}}$; a chronologically
         valid orientation; a target depth $D$
\Ensure time labels $\tau$ with $|\tau|\le D$, respecting
        Eq.~\eqref{eq: QT canonical orientation supp} at every data qubit, or
        \textsc{fail}
\State $E_{\rm free} \gets$ all edge classes of $T^{(\mathrm{base})}_{\mathrm{QT}}$
\For{$t = 1,\dots,D$}
  \State $\mathcal R_t \gets \bigl\{\, e \in E_{\rm free} \;:\;
         \nexists\, e' \in E_{\rm free} \text{ with } e' \prec e \,\bigr\}$
         \Comment{ready}
  \State $\mathcal K_t \gets \bigl\{\, v \;:\;
         \deg_{\rm free}(v) = D-t+1 \,\bigr\}$
         \Comment{critical}
  \State $\Gamma_t \gets$ a matching of $\mathcal R_t$ saturating
         $\mathcal K_t$, grown by augmenting paths
  \If{no such matching exists}
    \State \Return \textsc{fail}
  \EndIf
  \State extend $\Gamma_t$ to a maximal matching of $\mathcal R_t$, visiting the
         edges in decreasing order of $\deg_{\rm free}$ summed over their two
         endpoints
  \State $\tau(\Gamma_t) \gets t$; \quad
         $E_{\rm free} \gets E_{\rm free} \setminus \Gamma_t$
\EndFor
\State \Return $\tau$ if $E_{\rm free}=\emptyset$, and \textsc{fail} otherwise
\end{algorithmic}
\end{algbox*}

We can loosely estimate the upper bound for a schedule respecting Eq.~\eqref{eq: QT canonical orientation supp}. If we demand that each time band must be finished before the next time band, the naive upper bound is simply the sum of the chromatic number $\chi$ needed in each time band 
\begin{equation}
    \label{eq: QT crude upper bound supp}
    |\tau| \leq \chi(E) + \chi(M) + \chi(L). 
\end{equation}
We emphasize that this upper bound is loose, since Eq.~\eqref{eq: QT canonical orientation supp} only enforces the partial order on individual data qubits. This means that while one data qubit is still in the early time band, the others can already move to the middle time band. However, it can be a good indicator on how to divide the row and column sets into two halves. A heuristic approach is to choose the halving such that the RHS of Eq.~\eqref{eq: QT crude upper bound supp} is minimized. Depending on the code structures, an optimized halving can lead to the crude upper bound equals $\Delta$, and thus we can use the bound to check whether the schedule can obtain optimal depth. 

Nevertheless, due to the simple partial order given in Eq.~\eqref{eq: QT canonical orientation supp}, we can devise a simple greedy schedule algorithm where we build the circuit layer by layer with a targeted depth $D$ (see Box~\ref{alg: greedy schedule}). We write $E_{\rm free}$ for the edges of the base Tanner graph that have not been used in the previous time layers, $\deg_{\rm free}(v)$ for the degree of a vertex $v$ restricted to $E_{\rm free}$, and we extend the relation $\prec$ of Eq.~\eqref{eq: QT canonical orientation supp} from the check families to the edges themselves: two edges meeting the same data qubit inherit the order of their families, and edges on different data qubits are incomparable.
 
The main idea is that at any given time layer $t$, the edges that can
participate are exactly those with no predecessor left,
\begin{equation}
  \label{eq: QT ready set}
  \mathcal R_t = \bigl\{\, e \in E_{\rm free} \;:\;
                  \nexists\, e' \in E_{\rm free} \text{ with } e' \prec e \,\bigr\},
\end{equation}
since every edge that must precede such an $e$ has already been used. We also need to pay attention to the set $\mathcal K_t$ of vertices $v$ with $\deg_{\rm free}(v) = D-t+1$. We call these vertices critical, because their remaining edges must occupy every one of the $D-t+1$ layers that are left. As a result, we need a matching $\Gamma_t$ of $\mathcal R_t$ which saturates $\mathcal K_t$ at least. If additional spots are available, we add further edges
of $\mathcal R_t$ to $\Gamma_t$, where priority is given to edges with high-degree endpoints. We assign the time label $\tau(\Gamma_t) = t$ and remove $\Gamma_t$ from the available edge set $E_{\rm free}$ in subsequent layers. The algorithm runs until no edges are left in $E_{\rm free}$, or until no matching $\Gamma_t$ saturates $\mathcal K_t$; in the latter case we declare the search a failure and restart it, with a different tie-break or a larger $D$.

In Table~\ref{tab:qt-base} we solve the reduced problem for pairs of the local codes $[2,1,2]$, $[6,3,3]$ and $[8,4,4]$ used in Ref.~\cite{leverrier2025smallquantumtannercodes} to construct small QT codes, together with $[4,2,2]$. For each pair we sweep over a small subset of inequivalent column permutations of $H_1$ and $H'_1$: 30 for $[6,3,3]$ and $[8,4,4]$, 3 for $[4,2,2]$, and 1 for $[2,1,2]$. Since the reduced problem does not depend on the group $G$ or the multisets $\mathcal A,\mathcal B$, each row covers every QT code built from that pair. Restricted to the canonical orientation, the greedy algorithm of Box~\ref{alg: greedy schedule} reaches $\Delta$ on 3136 of the 3180 permutations. On the remaining 44 we use CP-SAT to decide whether any depth-$\Delta$ circuit respecting the orientation exists. For 24 of them none does for any choice of halving, so the miss can be due to the choice of orientation and not due to the greedy algorithm; the greedy falls short of an existing depth-$\Delta$ circuit on only 20 permutations. The left--right circuit of
Ref.~\cite{strikis2026syndromeextractioncircuits} is quoted by its amortized
depth, the depth per round averaged over many syndrome-extraction rounds, excluding measurement and initialization. Even so, it reaches $\Delta$ on only 240 of the 3180 permutations, specifically when the Tanner graph $T_{\rm QT}^{\rm (base)}$ is dense. In contrast, the greedy algorithm performs well on dense codes but can miss some circuits on more sparser codes.

\begin{table*}
\centering
\begin{tabular}{@{}ll rr rrrr rr@{}}
\toprule
\multicolumn{2}{c}{local codes} & \multicolumn{2}{c}{$T^{(\rm base)}_{\rm QT}$}
 & \multicolumn{4}{c}{Canonical orientation}
 & \multicolumn{2}{c}{left--right~[24]} \\
\cmidrule(r){1-2}\cmidrule(lr){3-4}\cmidrule(lr){5-8}\cmidrule(l){9-10}
$C_0,C_1$ & $C'_0,C'_1$ & $|E|$ & $\Delta$
 & total & greedy & CP-SAT only & $\Delta$ infeasible
 & $\bar T_{\rm LR}$ & at $\Delta$ \\
\midrule
$[2,1,2]$ & $[6,3,3]$ &  72 &  6 &  30 &  27 & 0 & 3 & 6--8   &  6 \\
$[4,2,2]$ & $[6,3,3]$ & 144 &  6 &  90 &  75 & 6 & 9 & 6--9   & 18 \\
$[6,3,3]$ & $[2,1,2]$ &  72 &  6 &  30 &  27 & 0 & 3 & 6--8   &  6 \\
$[6,3,3]$ & $[4,2,2]$ & 144 &  6 &  90 &  77 & 4 & 9 & 6--8   & 18 \\
\addlinespace[2pt]
$[2,1,2]$ & $[8,4,4]$ & 128 &  8 &  30 &  30 & 0 & 0 & 8--10  & 24 \\
$[4,2,2]$ & $[8,4,4]$ & 256 &  8 &  90 &  87 & 3 & 0 & 8--12  & 72 \\
$[8,4,4]$ & $[2,1,2]$ & 128 &  8 &  30 &  30 & 0 & 0 & 8--10  & 24 \\
$[8,4,4]$ & $[4,2,2]$ & 256 &  8 &  90 &  84 & 6 & 0 & 8--12  & 72 \\
\addlinespace[2pt]
$[6,3,3]$ & $[6,3,3]$ & 324 & 10 & 900 & 899 & 1 & 0 & 12--14 &  0 \\
\addlinespace[2pt]
$[6,3,3]$ & $[8,4,4]$ & 576 & 14 & 900 & 900 & 0 & 0 & 16--17 &  0 \\
$[8,4,4]$ & $[6,3,3]$ & 576 & 14 & 900 & 900 & 0 & 0 & 16--17 &  0 \\
\midrule
\multicolumn{4}{r}{total} & 3180 & 3136 & 20 & 24 & & 240 \\
\bottomrule
\end{tabular}
\caption{Parity-check circuits on the base graph $T^{(\rm base)}_{\rm QT}$ for each pair of local codes, swept over a small subset of inequivalent column permutations of $H_1$ and $H'_1$ (column ``total''). Under the canonical orientation, the next three columns partition these permutations: the greedy algorithm reaches $\Delta$; the greedy fails but CP-SAT reaches $\Delta$; or CP-SAT certifies that no circuit of depth $\Delta$ respects the orientation, for any choice of halving. For the left--right circuit of Ref.~\cite{strikis2026syndromeextractioncircuits}
we give the range of its amortized depth $\bar T_{\rm LR}$ and the number of permutations on which it reaches $\Delta$.}
\label{tab:qt-base}
\end{table*}

We also schedule the 27 QT codes whose parity-check matrices are provided in the auxiliary files of Ref.~\cite{leverrier2025smallquantumtannercodes}, spanning three pairs of local codes and a range of lift groups; Table~\ref{tab:qt-lift} lists their parameters and the depths we obtain. Recomputing $n$ and $k$ from the ranks of $H_X$ and $H_Z$ reproduces the reported values except for the $C_{21}$ instance, whose matrices give $k=4$ rather than $k=2$. The greedy algorithm reaches $\Delta$ on 25 of the 27 codes, at roughly half the depth of the alternating circuit. The two exceptions, lifted by $C_7$ and $C_{14}$, reach $\Delta+1$, and CP-SAT certifies that no depth-$\Delta$ circuit respecting the canonical orientation exists for either, for any choice of halving. These are not independent
failures: both codes have the same base graph and are thus consistent with with the three infeasible permutations of $[6,3,3]\times[2,1,2]$ in Table~\ref{tab:qt-base}.

\begin{table}
\centering
\renewcommand{\arraystretch}{1.15}
\begin{tabular}{@{}l @{\hspace{2.0em}} r @{\hspace{1.4em}} r @{\hspace{1.4em}} r @{\hspace{2.4em}} r@{}}
\toprule
$G$ & $n$ & $k$ & $d$ & depth \\
\midrule
\multicolumn{5}{@{}l}{$[6,3,3]\times[2,1,2]$, \ $w=6$, \ $\Delta=6$, \ $\Delta_X{+}\Delta_Z=12$}\\
\addlinespace[2pt]
$C_5$    &  60 & 2 &  8 & 6 \\
$C_6$    &  72 & 2 &  9 & 6 \\
$C_7$    &  84 & 2 & 10 & $7^\dagger$ \\
$C_9$    & 108 & 2 & 12 & 6 \\
$C_{10}$ & 120 & 2 & 12 & 6 \\
$C_{11}$ & 132 & 2 & 14 & 6 \\
$C_{14}$ & 168 & 2 & 16 & $7^\dagger$ \\
$C_{18}$ & 216 & 2 & 18 & 6 \\
$C_{21}$ & 252 & $4^\ddagger$ & 20 & 6 \\
\midrule
\multicolumn{5}{@{}l}{$[8,4,4]\times[2,1,2]$, \ $w=8$, \ $\Delta=8$, \ $\Delta_X{+}\Delta_Z=16$}\\
\addlinespace[2pt]
$C_5$    &  80 & 2 & 10 & 8 \\
$C_6$    &  96 & 2 & 12 & 8 \\
$C_8$    & 128 & 2 & 14 & 8 \\
$C_8$    & 128 & 6 & 12 & 8 \\
$C_{10}$ & 160 & 2 & 16 & 8 \\
$C_{12}$ & 192 & 2 & 18 & 8 \\
$C_{14}$ & 224 & 2 & 20 & 8 \\
$C_{16}$ & 256 & 2 & 22 & 8 \\
\midrule
\multicolumn{5}{@{}l}{$[6,3,3]\times[6,3,3]$, \ $w=9$, \ $\Delta=10$, \ $\Delta_X{+}\Delta_Z=18$}\\
\addlinespace[2pt]
$C_2\times C_2$  & 144 &  8 & 12 & 10 \\
$C_2\times C_2$  & 144 & 12 & 11 & 10 \\
$C_6$            & 216 &  8 & 18 & 10 \\
$Q_8$            & 288 & 16 & 16 & 10 \\
$C_8$            & 288 &  8 & 19 & 10 \\
$C_9$            & 324 &  4 & 26 & 10 \\
$C_{11}$         & 396 &  2 & 29 & 10 \\
$C_6\times C_2$  & 432 & 20 & $\leq22$ & 10 \\
$C_6\times C_2$  & 432 & 24 & 18 & 10 \\
$C_4\rtimes C_4$ & 576 & 28 & $\leq24$ & 10 \\
\bottomrule
\end{tabular}
\caption{Quantum Tanner codes from the auxiliary file of Ref.~\cite{leverrier2025smallquantumtannercodes}, grouped by their pair of local codes. For each code we list the lift group $G$, the parameters $[[n,k,d]]$ and the depth of our circuit; each block header gives the lower bound $\Delta$ and the depth $\Delta_X{+}\Delta_Z$ of the alternating circuit, both constant within the block. $^\dagger$~No depth-$\Delta$ circuit respects the canonical orientation, as certified by CP-SAT. $^\ddagger$~Computed from the supplied
matrices; Ref.~\cite{leverrier2025smallquantumtannercodes} reports $k=2$.}
\label{tab:qt-lift}
\end{table}

\end{document}